\documentclass[a4paper,11pt]{article} 
\pdfoutput=1

\usepackage{graphicx,rotating,hyperref,slashed,amsmath,amssymb,amsfonts,colortbl,cite,soul,cancel,cleveref,framed,ulem} 
\usepackage{graphicx} 
\usepackage{dsfont}
\usepackage{hyperref}
\usepackage[dvipsnames]{xcolor}

\usepackage[export]{adjustbox}
\makeatletter   
\hypersetup{colorlinks,bookmarksopen,bookmarksnumbered,
linkcolor=blue,pdfstartview=FitH,urlcolor=rossos,citecolor=verde}
\allowdisplaybreaks	 
\numberwithin{equation}{section} 
  
\hypersetup{%
    ,urlcolor=black
    ,citecolor=black
    ,linkcolor=black 
    }
\usepackage{mciteplus}

\AtBeginDocument{
  \hypersetup{
    urlcolor=blue,
    citecolor=blue,
    linkcolor=blue,
  }%
}
\interfootnotelinepenalty=\@M

\newcommand{\qev}[1]{\ensuremath{\langle {#1} \rangle}}
\newcommand{\be}{\begin{equation}}
	\newcommand{\ee}{\end{equation}}
\newcommand{\bea}{\begin{eqnarray}}
	\newcommand{\eea}{\end{eqnarray}}
\newcommand{\barr}{\begin{array}}
	\newcommand{\earr}{\end{array}}
\newcommand{\ba}{\begin{align}}
	\newcommand{\ea}{\end{align}}
\newcommand{\GW}{\ensuremath{\mathrm{GW}}}

\newcommand{\Omegagw}{\ensuremath{\Omega_{\rm GW}}}

\newcommand{\lc}{\ensuremath{\mathcal{L}}}
\newcommand{\oc}{\ensuremath{\mathcal{O}}}

\newcommand{\fc}{\ensuremath{\mathcal{F}}}

\newcommand{\phat}{\hat{p}}

\newcommand{\idm}{\ensuremath{\mathds{1}_M}}
\newcommand{\idn}{\ensuremath{\mathds{1}_N}}
\newcommand{\calm}{\ensuremath{\mathcal{M}}}

\def\bea{\begin{eqnarray}}
\def\eea{\end{eqnarray}} 

\def\be{\begin{equation}}
\def\ee{\end{equation}} 
\def\beq{\begin{equation}}
\def\eeq{\end{equation}}

\newcommand{\bx}{{\bf{x}}}

\newcommand\ees{\end{eqnarray}}
\newcommand\bees{\begin{eqnarray}}

\def\bea{\begin{eqnarray}}
\def\eea{\end{eqnarray}}

\def\Tr{{\rm Tr}}

\def\Tr{{\rm Tr}}

\def\0{{\boldsymbol 0}}

\def\cb{{\boldsymbol{C}}}
\def\eb{{\boldsymbol{E}}}
\def\hb{{\boldsymbol{H}}}
\def\nb{{\boldsymbol{N}}}

\def\phat{\hat{p}}

\def\lsim{\mathrel{\rlap{\lower3pt\hbox{\hskip0pt$\sim$}}
   \raise1pt\hbox{$<$}}}         
\def\gsim{\mathrel{\rlap{\lower4pt\hbox{\hskip1pt$\sim$}}
   \raise1pt\hbox{$>$}}}         

 \newcommand{\sfootnote}[1]{}
\definecolor{bluc}{cmyk}{1,1,0,0.1}
\definecolor{rossoCP3}{cmyk}{0,.88,.77,.40}
\definecolor{rosso}{cmyk}{0,1,1,0.4}
\definecolor{rossos}{cmyk}{0,1,1,0.55}
\definecolor{rossoc}{cmyk}{0,1,1,0.2}
\definecolor{verdes}{cmyk}{0.92,0,0.59,0.4}

\hypersetup{colorlinks, bookmarksopen, bookmarksnumbered,
citecolor=verdes, linkcolor=bluc, pdfstartview=FitH, urlcolor=rossos}

\usepackage{multicol}
\usepackage{color}
\definecolor{rosso}{cmyk}{0,1,1,0.4}
\definecolor{rossos}{cmyk}{0,1,1,0.55}
\definecolor{rossoc}{cmyk}{0,1,1,0.2}
\definecolor{blu}{cmyk}{1,1,0,0.3}
\definecolor{blus}{cmyk}{1,1,0,0.6}
\definecolor{bluc}{cmyk}{1,1,0,0.1}
\definecolor{verde}{cmyk}{0.92,0,0.59,0.25}
\definecolor{verdec}{cmyk}{0.92,0,0.59,0.15}
\definecolor{verdes}{cmyk}{0.92,0,0.59,0.4}

\skip\@mpfootins = \skip\footins
\renewcommand\&{&}

\def\circa#1{\,\raise.3ex\hbox{$#1$\kern-.75em\lower1ex\hbox{$\sim$}}\,}

\newfam\rsfsfam
\def\mathscr#1{{\fam\rsfsfam\relax#1}}

\def\circa#1{\,\raise.3ex\hbox{$#1$\kern-.75em\lower1ex\hbox{$\sim$}}\,}
\makeatletter

\def\hhref#1{\href{http://arxiv.org/abs/#1}{arXiv:#1}} 

\newcommand{\doi}[1]{\href{http://dx.doi.org/#1}{[doi]}}

\def\hhref#1{\href{http://arxiv.org/abs/#1}{arXiv:#1}} 
 
\def\art{\@ifnextchar[{\eart}{\oart}}
\def\eart[#1]#2#3#4#5#6{{\rm #2}, {\em #3 \bf #4} {\rm (#6) #5} ({\em #1})}

\def\article{\@ifnextchar[{\earticle}{\oarticle}}
\def\oarticle#1#2#3#4#5#6{{\rm #1}, {\em ``#6''}, {\rm #2 #3 (#5) #4}}
\def\earticle[#1]#2#3#4#5#6#7{{\rm #2}, {\em ``#7''}, {\rm #3 #4 (#6) #5}  [\hhref{#1}]}
\def\hepart[#1]#2{{\rm #2, \em#1}}
\def\heparticle[#1]#2#3{#2, {\em ``#3''} [\hhref{#1}]}

\newcounter{alphaequation}[equation]
\def\thealphaequation{\theequation\hbox to
0.6em{\hfil\alph{alphaequation}\hfil}}
\def\eqnsystem#1{
\def\@eqnnum{{\rm (\thealphaequation)}}
\def\@@eqncr{\let\@tempa\relax \ifcase\@eqcnt \def\@tempa{& & &} \or
  \def\@tempa{& &}\or \def\@tempa{&}\fi\@tempa
  \if@eqnsw\@eqnnum\refstepcounter{alphaequation}\fi
\global\@eqnswtrue\global\@eqcnt=0\cr}
\refstepcounter{equation} \let\@currentlabel\theequation \def\@tempb{#1}
\ifx\@tempb\empty\else\label{#1}\fi
\refstepcounter{alphaequation}
\let\@currentlabel\thealphaequation
\global\@eqnswtrue\global\@eqcnt=0 \tabskip\@centering\let\\=\@eqncr
$$\halign to \displaywidth\bgroup \@eqnsel\hskip\@centering
$\displaystyle\tabskip\z@{##}$&\global\@eqcnt\@ne
\hskip2\arraycolsep\hfil${##}$\hfil& \global\@eqcnt\tw@\hskip2\arraycolsep
$\displaystyle\tabskip\z@{##}$\hfil
\tabskip\@centering&\llap{##}\tabskip\z@\cr}
\def\endeqnsystem{\@@eqncr\egroup$$\global\@ignoretrue} \makeatother

\definecolor{fiorentina}{rgb}{.5,0,.5}

\newcommand {\bn}[0]{{\bf n}}
\newcommand {\bp}[0]{{\bf p}}
\newcommand {\bv}[0]{{\bf v}}
\newcommand {\bq}[0]{{\bf q}}

\begin{document}

\setcounter{page}{1} \baselineskip=15.5pt \thispagestyle{empty}

\vspace{0.8cm}
\begin{center}

{\fontsize{19}{28}\selectfont \sffamily \bfseries   {{  Astrometry
meets 
Pulsar Timing Arrays:}
\\
\vskip0.2cm
 Synergies for Gravitational Wave Detection
}}

\vspace{0.5cm}

\begin{center}
{\fontsize{12}{30}\selectfont  
N.~M.~Jim\'enez Cruz$^{a}$ \footnote{\texttt{nmjc1209.at.gmail.com}}, Ameek Malhotra$^{a}$ \footnote{\texttt{ameek.malhotra.at.swansea.ac.uk}}, Gianmassimo Tasinato$^{a, b}$ \footnote{\texttt{g.tasinato2208.at.gmail.com}}, Ivonne Zavala$^{a}$ \footnote{\texttt{e.i.zavalacarrasco.at.swansea.ac.uk}}
} 
\end{center}

\begin{center}

\vskip 8pt
  \textsl{$^{a}$ Physics Department, Swansea University, SA28PP, United Kingdom}\\
\textsl{$^{b}$ Dipartimento di Fisica e Astronomia, Universit\`a di Bologna,\\
 INFN, Sezione di Bologna, I.S. FLAG, viale B. Pichat 6/2, 40127 Bologna,   Italy}
\vskip 7pt

\end{center}

\smallskip
\begin{abstract}
\noindent
High-precision astrometry offers a promising approach to detect low-frequency gravitational waves, complementing pulsar timing array (PTA) observations. We explore the response of astrometric measurements to a stochastic gravitational wave background (SGWB) in synergy with PTA data. Analytical, covariant expressions for this response are derived, accounting for the presence of a possible dipolar anisotropy in the SGWB. 
We identify the optimal estimator for extracting SGWB information from astrometric 
observations and examine how sensitivity to SGWB properties varies with the sky positions of stars and pulsars. Using representative examples of current PTA capabilities and near-future astrometric sensitivity, we demonstrate that cross-correlating astrometric and PTA data can improve constraints on SGWB properties, compared to PTA data alone.
The improvement is quantified through Fisher forecasts for the SGWB amplitude, spectral tilt, and dipolar anisotropy amplitude. In the future, such joint constraints could play a crucial role in identifying the origin of SGWB signals detected by PTAs.
\end{abstract}

\end{center}

\section{Introduction}


The detection of a stochastic background of gravitational waves (SGWB) would represent a major milestone for gravitational wave astronomy. Recent exciting developments in the nano-Hertz (nHz) frequency regime suggest that we may be on the verge of  a detection, with several Pulsar Timing Array (PTA) collaborations reporting strong evidence for a SGWB in this range~\cite{NANOGrav:2020bcs,Goncharov:2021oub,EPTA:2021crs,Antoniadis:2022pcn,NANOGrav:2023gor,EPTA:2023fyk,Reardon:2023gzh,Xu:2023wog,InternationalPulsarTimingArray:2023mzf,Miles:2024seg}.  Hellings-Downs (HD) correlations~\cite{Hellings:1983fr}, representing  the `smoking-gun' signature of SGWB, have been detected with $2-4\sigma$ of statistical significance, depending on the dataset. Upcoming data from PTAs, in particular the joint analysis from IPTA Data Release 3, are expected to further increase the significance of this detection and measure the SGWB amplitude and spectrum more precisely~\cite{InternationalPulsarTimingArray:2023mzf}. The observed signal may arise from mergers of supermassive black hole binaries (SMBHB)~\cite{Sesana:2008mz,Burke-Spolaor:2018bvk} or from early-universe sources~\cite{Caprini:2018mtu} (or a combination of both) and the increasing precision offered by future data may help in identifying the origin of the SGWB. In the longer term, the Square Kilometre Array (SKA) \cite{Janssen:2014dka,Keane:2014vja}, with its unparalleled timing precision and larger number of observed pulsars, will provide definitive measurements of the SGWB properties.

It is worthwhile to also explore alternative probes that could complement the PTA observations in the nHz regime. One such promising probe is astrometry -- using the precise measurements of the positions of a very large number of distant sources to measure the effects of gravitational waves~\cite{Linder_Astro,Braginsky:1989pv,Fakir_astro:1994,Pyne:1995iy,Kaiser:1996wk,Gwinn:1996gv,Kopeikin:1998ts,Schutz_2009, Book:2010pf,Moore:2017ity,Klioner:2017asb,Mihaylov:2018uqm,Qin:2018yhy,Mentasti:2023gmr,Inomata:2024kzr,An:2024axz,Jaffe:2004it,Lu:2024yuo,Zwick:2024hag}. GW present between the Earth and the source induce deflections in the observed positions of these sources (hereafter we lump all such sources under the umbrella term `stars'). Much like the pulsar timing residuals, which are correlated across different pulsars as a function of their angular separation, the astrometric deflections for different stars are also correlated, albeit with a slightly different form of the correlation as compared to the PTA HD correlation. 

The success of the Gaia mission~\cite{Gaia:2016zol,Gaia2023}, which observes billions of stars with a  precision measurement of their position   of the order of the milli-arcsecond (mas), has spurred a renewed interest in astrometric detection of SGWB. Previously, data from Very-long baseline interferometry (VLBI)~\cite{VLBI} and more recently Gaia, have already been used to set upper limits on the SGWB in the frequency range $10^{-18}\,\mathrm{Hz} \lesssim f \lesssim 10^{-9}\,\mathrm{Hz}$~\cite{Gwinn:1996gv,Titov_2011,Darling:2018hmc,Aoyama:2021xhj,Jaraba:2023djs,darling2024newapproachlowfrequency}. The full Gaia data release 5 (DR5) with the individual time-series measurements will further extend this frequency range to $f \lesssim 4\times 10^{-7} \mathrm{Hz}$ (each star is observed roughly 14 times a year~\cite{Gaia:2016zol}). The upcoming Roman survey~\cite{sanderson2019astrometrywidefieldinfraredspace} --  due to its much higher observing cadence --  may even be able to push towards higher frequencies ($f\lesssim 10^{-4}\,\rm Hz$), opening up an additional observational window for GW~\cite{Wang:2020pmf,Wang:2022sxn,Pardo:2023cag}. Although current astrometric upper limits on the SGWB lie around the $\Omegagw \lesssim 10^{-2}$ level, the proposed mission Theia~\cite{Theia:2017xtk,Malbet:2022lll}, with its $\mu\rm as$ astrometric precision would  represent a significant upgrade in terms of sensitivity to SGWB~\cite{Garcia-Bellido:2021zgu,Jaraba:2023djs,An:2024axz}, even being competitive with PTA sensitivity. 

Motivated by these considerations, in this work we investigate how data from astrometric surveys may be used to complement existing PTA experiments,  and potentially improve upon current SGWB measurements obtained from PTA data alone.
We do this by means of a Fisher matrix analysis and calculate the improvement in constraints that can be achieved by cross-correlating timing residuals of PTAs with the angular deflections of astrometry, focusing on the parameters corresponding to the SGWB monopole amplitude, spectral tilt (for a power-law spectrum) and a possible  dipole anisotropy magnitude.  Along the way, we derive for the first time  analytic covariant expressions for the response to dipole anisotropy in the SGWB, for astrometry as well as its cross-correlation with PTAs. The anisotropy of the nHz SGWB is a key observable that may be used to discriminate between astrophysical and cosmological origins of the SGWB given their differing predictions, with the magnitude of  intrinsic cosmological SGWB anisotropies (see e.g.~\cite{Alba:2015cms,Contaldi:2016koz,Geller:2018mwu,Bartolo:2019oiq,Bartolo:2019zvb,Bartolo:2019yeu,Dimastrogiovanni:2021mfs,LISACosmologyWorkingGroup:2022kbp,Adshead:2020bji}) expected to be much smaller than their astrophysical counterpart~\cite{Mingarelli:2013dsa,Taylor:2013esa,Mingarelli:2017PTA,Taylor:2020zpk,Gardiner:2023zzr,Sato-Polito:2023spo,NANOGrav:2023tcn,Sah:2024oyg,Raidal:2024tui}. Cosmological
anisotropies of large magnitude include {\it kinematic anisotropies} due to our motion
with respect to the primordial source of  SGWB \cite{Cusin:2022cbb,LISACosmologyWorkingGroup:2022kbp,Chowdhury:2022pnv,Tasinato:2023zcg,Bertacca:2019fnt,ValbusaDallArmi:2022htu,Chung:2022xhv}. For primordial sources of SGWB, we
can expect a dipolar anisotropy with an amplitude one thousand smaller than the isotropic part of the background, 
as found in cosmic microwave background measurements \cite{Smoot:1977bs,Kogut:1993ag,WMAP:2003ivt,Planck:2013kqc}.
Given that kinematic anisotropies of the SGWB are well motived, we mainly consider a kinematic dipole as specific target for our analysis, although it can be easily extended to discuss other anisotropies as well.
Additional works 
 discussing
detection prospects of  the
anisotropies of the SGWB   in different
contexts are
\cite{Allen:1996gp,Anholm:2008wy,Mingarelli:2013dsa,Taylor:2013esa,Gair:2014rwa,Hotinli:2019tpc,Ali-Haimoud:2020iyz,Ali-Haimoud:2020ozu,LISACosmologyWorkingGroup:2022jok} with PTA specific analyses in~\cite{Cornish:2014rva,Taylor:2015udp,NANOGrav:2023tcn,Depta:2024ykq,Konstandin:2024fyo,Pol:2022sjn,Lemke:2024cdu}. See e.g.~\cite{Bernardo:2024bdc} for a review.

This paper is organised as follows: in \cref{sec:theory_ORF} we review the theory of SGWB detection with astrometry and corresponding monopole response and overlap reduction functions. We then derive an analytic expression for the astrometric response to the kinematic dipole anisotropy and its auto-correlation. We also discuss the astrometry-redshift correlation for the SGWB monopole and derive an analytic expression for this cross-correlation in the presence of a dipole anisotropy.  In \cref{sec:forecasts}, we use these results to forecast the astrometric sensitivity to the SGWB monopole and dipole and also estimate the improvement in constraints that can be obtained by cross-correlating astrometry and PTA data, over PTA data alone. Finally, in \cref{sec:conclusions}, we present our conclusions. Two technical appendixes
complement our discussion.

\section{Overlap  functions for astrometry and pulsar timing arrays}
\label{sec:theory_ORF}

{The overlap
reduction function (ORF) of a pair of detectors accounts for how the sensitivity to the SGWB changes due to the relative positions and orientations of the detectors. The cross-correlation of the signal across the two detectors can be written entirely in terms of the ORF and the underlying SGWB spectral density. Schematically, one can write 
\begin{align}
    \qev{s_I(f) s_J(f) } \propto \gamma_{IJ}(f)I(f)\,,
\end{align}
where $s_{I,J}$ denotes the signal at detector $I,J$, $\gamma_{IJ}$ the overlap reduction function and $I(f)$ the SGWB intensity.} In this  section we obtain  expressions
for the astrometric ORFs and their cross-correlations with pulsar timing arrays.  We include the possible presence of a dipolar SGWB anisotropy, which can be
instrumental for distinguishing cosmological from astrophysical backgrounds. We investigate the crucial role
of the position of the monitored objects for determining the sensitivity of the system to the SGWB properties. Our covariant, analytical formulas are convenient and simple to use for Fisher
forecast analysis.

Astrometry aims to measure the position and motion
of stars on the celestial sphere,
while
PTA are sensitive to the time of arrival of radio signals from distant
pulsar sources.  GW  passing between the
position of 
stars, pulsars, 
and the Earth  modify photon geodesics, leading to   effects which are measurable 
 with both methods. 
 The presence of a stochastic
 GW background (SGWB) induces correlations among measurements  of GW signals, as detected by observations of distinct astronomical objects. In most studies the SGWB is assumed isotropic.
  However, it  develops kinematic anisotropies  if the frames of GW source and GW detector move  with respect to each other \cite{Cusin:2022cbb}. Kinematic anisotropies break
 the isotropy of the SGWB, making
 it direction-dependent:  they are  fully determined by the properties
 of the isotropic part
 of the background, as well as the direction of the velocity among frames.  For this reason it is  desirable  to derive
 analytical, covariant expressions
 for the ORF, which make transparent
 how the sensitivity to  
SGWB properties depend on the position of stars in the sky.
We do so
 in section \ref{sec:astrov} where we focus specifically on astrometry, while in section \ref{sec_xcorr} 
we analyse its possible synergies with PTA observations.

\subsection{Astrometry overlap
functions}\label{sec:astrov}

We define 
GW in terms of spin-2 fluctuations of the flat metric
\be
d s^2 \,=\,- dt^2+\left[ \delta_{ij}+h_{ij}(t,\vec x) \right]\,d x^i d x^j
\,.
\ee
The tranverse-traceless spin-2 tensors $h_{ij}(t,\vec x)$ can be decomposed in Fourier
modes as 
\be
\label{fouhij}
h_{ij}(t,\vec x)\,=\,\sum_{\lambda}\,\int_{-\infty}^{+\infty} d f\,\int d^2 \bp\,{ e}^{-2 \pi i f \,\bp \cdot \vec x}\,e^{2 \pi i f t}\,
{\bf e}_{ij}^\lambda (\bp)\,h_{\lambda}(f, \bp)\,,
\ee
where $f$ is the GW frequency, and $\bp$ the unit vector indicating 
its direction.
  ${\bf e}_{ij}^{\lambda}(\bp)$  are real polarization tensors in the basis  $\lambda=(+,\times)$. {Notice that we take the flat space limit of the GW evolution, hence it satisfies a  flat space d’Alembertian equation.} We impose $h_\lambda (-f, \bp)\,=\,h^*_\lambda (f, \bp)
$. A  GW passing between a star and the Earth causes a distortion $\delta n^i$
on the star direction, $\bn$, at time $t$, given by the formula (see e.g. \cite{Book:2010pf}) 
\be
\label{expdef}
\delta n_i(t, \bn)\,=\,{\cal R}_{ikl} (\bn, \bp)\,h_{kl}(t,0)\,,
\ee
where as customary we focus on the so-called  `Earth-term' contribution only (see e.g. \cite{Book:2010pf,Maggiore:2018sht}). 
We introduce 
\be \label{eq:R_ikl}
{\cal R}_{ikl} (\bn, \bp)\,=\,\frac{n_k }{2} \left[\frac{(n_i+p_i) n_l}{1+\bn \cdot \bp}
-\delta_{il}\,,
 \right]
\ee
where recall that the unit vectors
$\bn$ and $\bp$ indicate the direction of the star and of GW propagation, respectively. The three-index quantity ${\cal R}_{ikl}$ is orthogonal to the unit vector  $n_i$
when contracted with  its first index. 

In order to characterise {the stochastic GW signal, we assume it to have a Gaussian distribution with zero mean.}. {Thus, the SGWB can be entirely characterised by its two-point function, expressed as}
\be
\label{htpf}
\langle h_{\lambda_1}(f_1, \bp_1)
h^*_{\lambda_2}(f_2, \bp_2)
\rangle
\,=\,
\frac{3 H_0^2}{4 \pi^2}\,\frac{\Omega_{\rm GW} (f_1, \bp_1)}{f_1^3}\,\delta_{\lambda_1 \lambda_2}
\,\delta(f_1-f_2)\,\frac{
\delta^{(2)}(\bp_1-\bp_2)
}{4\pi}\,,
\ee
where {the angle brackets denote an ensemble average, over all possible realisations of the SGWB} and $\Omega_{\rm GW}(f, \bp)$ is the  quantity   customarily used for characterizing
the GW energy density per log frequency interval (see e.g. \cite{Maggiore:2007ulw}). The spectral energy density parameter $\Omegagw$ is related to the GW intensity, $I$, through the relation
    \begin{align}
    \label{rel_ogwi}
        \Omegagw(f, \bp)  = \frac{4\pi^2 f^3}{3H_0^2}I(f, \bp) \,.
    \end{align}
A SGWB can be characterized  by  anisotropies, which
render  $ \Omega_{\rm GW}(f, \bp)$ explicitly
dependent on the GW direction $\bp$. We include here
 SGWB {\it kinematic anisotropies} \cite{Cusin:2022cbb}, induced
by the motion of the observer with respect to the GW background source. 
They can be a key observable
for distinguishing astrophysical
from cosmological
backgrounds \cite{Cusin:2022cbb,LISACosmologyWorkingGroup:2022kbp}. Denoting with  
{$\beta = |\bv|/c$} the size
of the relative velocity among frames with
respect to the  speed of light, and $\bv$ the velocity
unit vector, we have \cite{Cusin:2022cbb}
\be
\label{kinan2}
\Omega_{\rm GW}(f, \bp)\,=\,\bar \Omega_{\rm GW} (f)+\beta (4-n_\Omega)
\,\bp \cdot \bv\,\bar \Omega_{\rm GW} (f)\,,
\ee
where $\bar \Omega_{\rm GW} (f) = 
(4 \pi)^{-1}\int d^2 \bp \,\Omega_{\rm GW}(f, \bp)$ is the  angular  averaged GW energy density.
The result depends
also on the energy density spectral
tilt $n_\Omega\,\equiv\,d \ln \Omega_{\rm GW}/d \ln f$.
The  simple form of the SGWB anisotropy of 
equation \eqref{kinan2} is all what we
need to carry on our calculations. We
include
only the kinematic
dipole contribution 
proportional
to $\beta$, working under
the hypothesis of small relative
 velocity among frames --  supported by  observations of the cosmic
microwave background (CMB) dipolar
anisotropy of the CMB temperature \cite{Smoot:1977bs,Kogut:1993ag,WMAP:2003ivt,Planck:2013kqc}.
 It is important to emphasize that kinematic effects
 are associated with  deterministic  -- and not statistical -- anisotropies. Hence, formula
 \eqref{kinan2} fully 
 characterizes the kinematic
 dipolar anisotropy, with no need
 of ensemble averaging. An implication is that our analysis of overlap functions
 can be conveniently carried
 on in a fully covariant way.

An alternative method would be to work in the spherical harmonic basis for the timing residuals and angular deflections, e.g. see in~\cite{Roebber:2016jzl,Qin:2018yhy,Nay:2023pwu,Caliskan:2023cqm,Inomata:2024kzr}. For a full-sky survey with a uniform distribution of stars, spherical harmonics provide a convenient diagonal basis for analysing the effects of the SGWB due to the orthogonality of the spherical harmonics. However, in general the dataset used may not have a uniform distribution of sources or be full-sky e.g. see~\cite{Jaraba:2023djs}. To keep our analysis completely general and applicable to arbitrary distributions of stars, we do not resort to the spherical harmonic decomposition. 

\smallskip

We proceed considering two-point correlators of
star deflections as induced by GW passing between stars and the Earth.   Denoting with $\bn$ and $\bq$ the unperturbed  directions of the two stars respectively, we use 
eqs \eqref{expdef}, \eqref{htpf} and \eqref{kinan2} to find
\bea
\langle \delta n^i(\bn, t)
 \delta n^j(\bq, t')
 \rangle&=&\frac{3 H_0^2}{32 \pi^3}
 \int df\,\frac{\bar \Omega_{\rm GW}(f)}{f^3}{\cos[2\pi f(t-t')]}
 \left[ H^{(0)}_{ij} (\bn, \bq)+ \beta\,(4-n_\Omega)\,
  H^{(1)}_{ij} (\bn, \bq, {\bf v})
 \right]\,, \nonumber
 \\
 &=&{p}_{(0)}\,H^{(0)}_{ij} (\bn, \bq)+
 {p}_{(1)}\,H^{(1)}_{ij} (\bn, \bq, \bv)
 \label{eq_secexp}\,.
\eea
The quantities ${p}_{(0,1)}$
depend on integrals
along the frequency of the 
SGWB amplitude and its spectral
tilt $n_\Omega$.
The tensors $H^{(0)}_{ij}$ and $H^{(1)}_{ij} $ are independent
of frequency.  
They are
formally expressed in terms of the following angular
integrals
\bea
\label{defh0}
 H^{(0)}_{ij} (\bn, \bq)
 &=& \int d^2 \Omega_{\bp}\,{\cal R}_{ikl} (\bn, \bp) {\cal R}_{jrs} (\bq, \bp)\,
 P_{kl rs}\,,
 \\
 \label{defh1}
 H^{(1)}_{ij} (\bn, \bq, \bv)
 &=& \int d^2 \Omega_{\bp}\, ( \bp
 \cdot
 {\bf v})
  \,{\cal R}_{ikl} (\bn, \bp) {\cal R}_{jrs} (\bq, \bp)\,
 P_{kl rs}\,.
\eea
with {the projector
\begin{eqnarray}
    \mathcal{R}_{ijk}(\mathbf{n}, \mathbf{p}) = 
    \frac{n_{j}}{2}\left[\frac{(n_{i}+p_i)\,n_{k}}{1+\mathbf{n}\cdot \mathbf{p}} - \delta_{ik}\right].
\end{eqnarray}
arising from properties of the polarization tensors
in this context \cite{Book:2010pf}.}
The tensors $H^{(0,1)}_{ij}$ are the analog
of the PTA ORF
relative to the monopolar
and  dipolar kinematic components
of the SGWB. 
To compute these quantities,
we introduce the  projection tensor 
\be \label{eq:project_Tensor}
P_{ijkl}
\,=\,\delta_{ik} \delta_{jl}
+\delta_{il} \delta_{jk}-\delta_{ij} \delta_{kl}
+p_i p_j p_k p_l
-\delta_{ik} p_j p_l 
-\delta_{jl} p_i p_k 
-\delta_{il} p_j p_k-\delta_{jk} p_i p_l
+\delta_{ij} p_k p_l+\delta_{kl} p_i p_j
\ee
associated with combinations
of polarization tensors $e_{ij}^{(\lambda)}$.
The tensors $H^{(0,1)}_{ij}(\bn, \bq)$ are orthogonal
to $\bn$ in their first index, and to $\bq$ in their second index. 

The integrals \eqref{defh0}, \eqref{defh1}, can
be performed  with a standard
methods of contour integration used  in the context
of PTA physics, see e.g. \cite{Jenet:2014bea}~\footnote{ { In fact, the calculation proceeds exactly as in \cite{Jenet:2014bea}. 
We decompose the angular integration measure as 
$d^2\Omega_{\mathbf{p}} = d(\cos\theta)\, d\varphi$, 
and perform the substitution $\varphi = e^{i z}$. 
In this way, the resulting integral in the complex plane is expressed 
as a contour integral over $\mathcal{C}$, corresponding to the unit circle. 
The integrand has poles inside the unit circle, and the 
integral and can be straightforwardly evaluated using Cauchy’s integral formula and the 
residue theorem.}}. For the case
of the monopole, the ORF
 is associated with the integral in eq \eqref{defh0}. We consider  the
 combination 
\bea
\label{defoy}
y&=&\frac{1-\bn \cdot \bq}{2}\,=\,\frac{1-\cos{\zeta}}{2}
\,
\eea
controlling
the angular separation between a pair of
stars on the celestial sphere: 
this quantity lies
in the interval $0\le y\le1$. The astrometry ORF relative to the monopolar, isotropic component of the SGWB is 
\bea
H^{(0)}_{ij} (\bn, \bq)&=&\frac{\pi}{3
(1-y)^2}
\left(1-8 y+7 y^2-6 y^2 \ln y 
\right)\times
\nonumber
\\
&&\left[ (2-2y)
\delta_{ij}
-n_i n_j-q_i q_j-q_i n_j+(1-2 y) q_j n_i
\right] \,,
\label{eq:H0ij}
\eea
a result equivalent to what was found
in \cite{Book:2010pf}, although expressed
slightly more compactly. As mentioned
above, we obtained
this expression by evaluating
the integral \eqref{defh0} using contour integration.
The expression
\eqref{eq:H0ij} is the astrometry equivalent of the
Hellings-Downs function, the PTA ORF for the SGWB
monopole. {From the statistical properties of the two-point correlators, one expects a dependence only on the angular separation $\bn \cdot \bq$ between stars in \eqref{eq:H0ij}. This feature is not immediately apparent in our results because of the tensorial structure introduced by the projection tensor \eqref{eq:project_Tensor}, which brings explicit factors of $n_i$ and $q_j$ into the expression. A useful way to reveal this property is through the quantity $\Tr[\hb_{0}\hb_{0}]$\footnote{{This expression also plays an important role in the Fisher forecasts of \cref{sec:forecasts}}.}, which, as expected, depends only on the relative directions (see Appendix~\ref{Trace_Appendix}). Alternative formulations that make the exclusive dependence on $\bn \cdot \bq$ manifest have also been explored; see, e.g., \cite{Mihaylov:2018uqm,Inomata:2024kzr}.}

The tensor defined by the integral \eqref{defh1} corresponds instead to
the astrometry response function
for dipolar kinematic  anisotropies. It can
be computed in terms of appropriate spherical
harmonics \cite{Inomata:2024kzr} -- but  we provide here
a covariant,  succinct  expression for it.
We introduce, as in \cite{Book:2010pf}, a  basis
of vectors:
\bea
{\bf A}&=& {\bf n} \times {\bf q}
\hskip0.5cm,\hskip0.5cm
{\bf B}\,=\, {\bf n} \times {\bf A}
\hskip0.5cm,\hskip0.5cm
{\bf C}\,=\, -{\bf q} \times {\bf A}\,.
\eea
Computing the integral 
of eq \eqref{defh1} by means 
of contour integrations,
we find
\be
H_{ij}^{(1)}\,=\,a_1 \,\left( A_i C_j+ \,B_i A_j\right)
+a_2 \left(B_i C_j- A_i A_j\right)
\label{eq:H1ij}\,,
\ee
with $a_{1,2}$ scalar coefficients depending
on the angles among the
vectors involved.  Recalling
the definition \eqref{defoy}, and denoting
\be
(Av)\,=\,\bv \cdot \left(\bn \times \bq \right)\hskip0.5cm, \hskip0.5cm
(nv)\,=\,\bn \cdot \bv\,,
\ee
we  express the coefficients
$a_{1,2}$
as

\bea
a_1&=&\frac{ \pi  ({Av})  \left(
1-4 y
-
\frac{3 y^2 \ln (y)}{1-y }\right)}{6\,y\,(1-y)^2}\label{eq:defa1}\,,
\\
a_2&=&
\frac{ \pi   \left[({y}-1) (2 {y}+1)-3 {y} \ln ({y})\right]\left[2 ({nv}) (1-{y})+\sqrt{4 (1-{nv})^2 (1-{y}) {y}-({Av})^2}\right]}{6  (1-{y})^3}\,.
\nonumber\\
\eea

\medskip

Our analytical, covariant expression
\eqref{eq:H1ij} 
for the  ORF demonstrates that the astrometry sensitivity to the SGWB {dipole}
depends not only on the angle among
stars, $\zeta$, but also on their position
in the sky, and on the velocity vector
$\bv$ among frames.   

\begin{figure}[h!]
    \centering              \includegraphics[width=0.45 \linewidth]{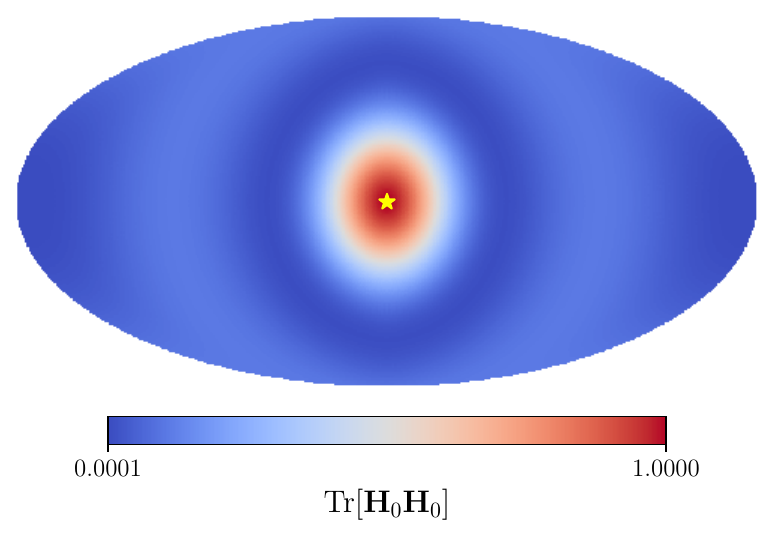}   \includegraphics[width=0.45 \linewidth]{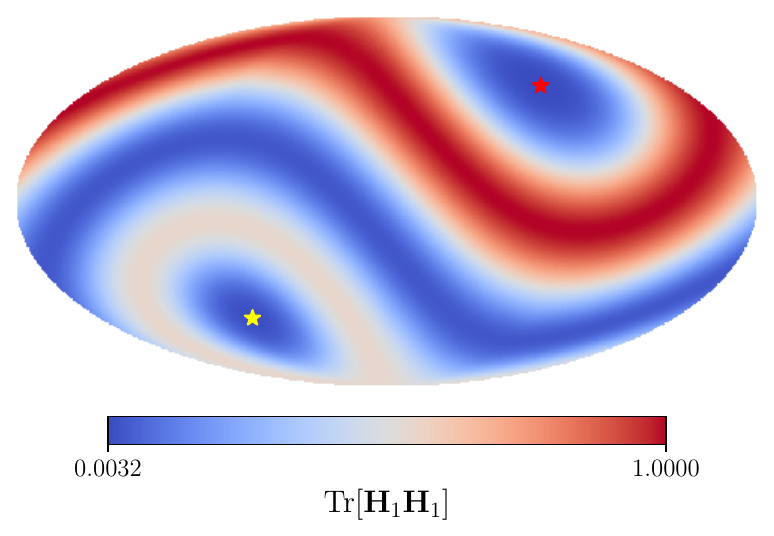}
    \includegraphics[width=0.45 \linewidth]{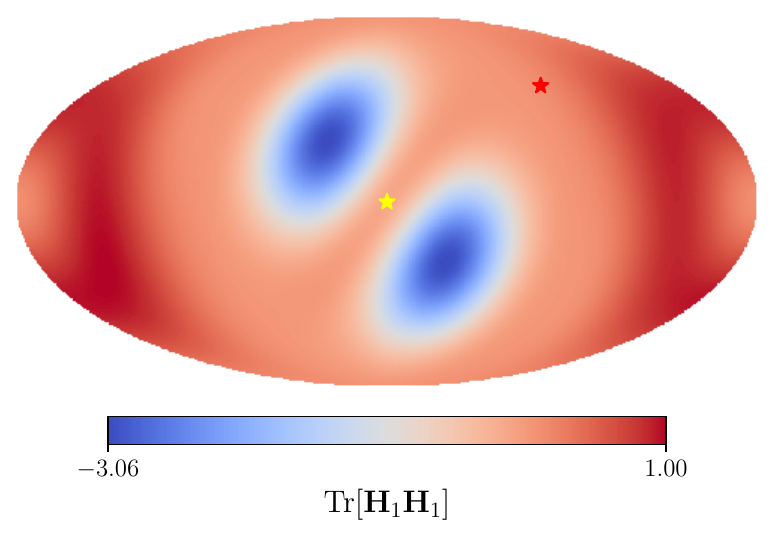}
    \includegraphics[width=0.45 \linewidth]{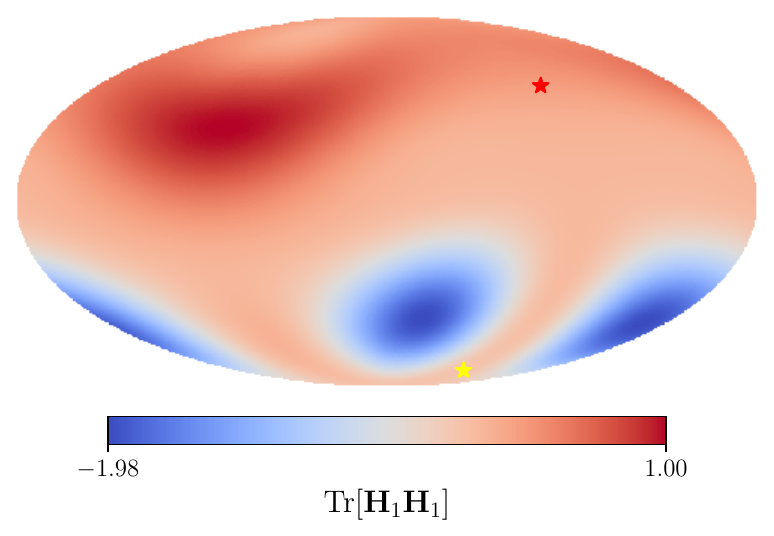}
    \caption{  \small The quantities $\Tr[\hb_{0}\hb_{0}]$ and $\Tr[\hb_{1}\hb_{1}]$, associated with response  to a SGWB  depending on stars positions. (See the main text for our notation.) Motivated by CMB, we choose  the kinematic dipole direction $\bv$ (red star) at $(l,b)=(264^{\circ},48^{\circ})$ in galactic coordinates.  Each panel shows a different choice of $\bn$, while the stars $\bq$ take the position of each pixel of the map.
  {\bf Upper left panel}: $\bn$ pointing towards $(l,b)=(0,0)$. 
 {\bf Upper right panel}: $\bn$ pointing towards $-\bv$. {\bf Lower left panel}:  $\bn$ at $(l,b)=(0,0)$. {\bf Lower right panel}: $\bn$ pointing towards the  direction  $(l,b)=(270.21^{\circ},-75.45^{\circ})$.
    }
    \label{Maps_H}
\end{figure}

The compact expressions 
\eqref{eq:H0ij} and
\eqref{eq:H1ij}
are particularly suitable to visualise patterns of sensitivity on the celestial sphere, 
and to further explore geometrical
features of physically relevant
quantities in the context we are examining. We plot  in  Figure \ref{Maps_H}
the combinations.

\begin{figure}[h!]
    \centering            \includegraphics[width=0.45 \linewidth]{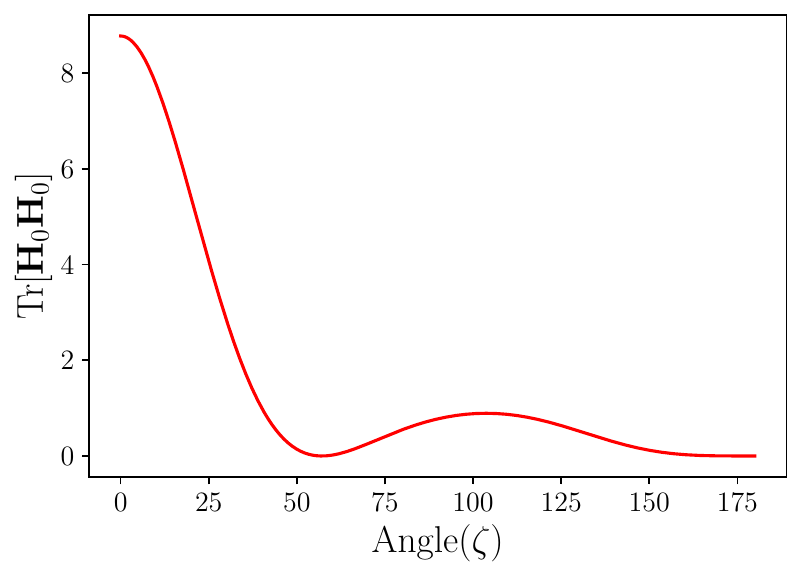}   \includegraphics[width=0.45 \linewidth]{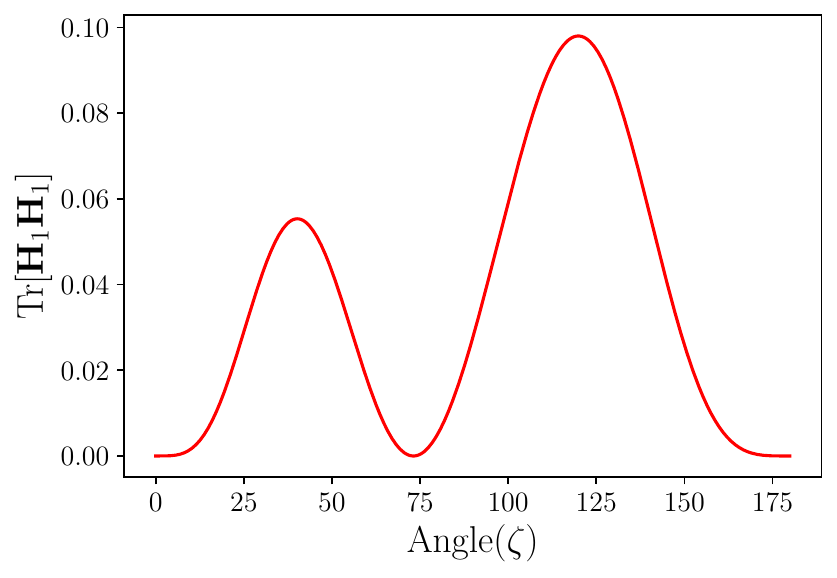}
    \caption{ \small Overlap functions response in terms of the angle between the stars. {\bf Left:} Angular dependence for
$\Tr[\hb_{0}\hb_{0}]$. {\bf Right:}  $\Tr[\hb_{1}\hb_{1}]$ when choosing $\bn = - \bv$.}
    \label{Plots_H}
\end{figure}

 $\Tr[\hb_{0}\hb_{0}]$ and $\Tr[\hb_{1}\hb_{1}]$, representative
of the sensitivity of astrometry observations
to the monopole (through the function $\hb_{0}$)
and the dipole (through the function $\hb_{1}$).\footnote{A dipolar anisotropy induces also a correlation between
the electric $E$ and magnetic $B$ components of the angular deflection correlation functions. We explore
this topic in Appendix \ref{sec_spec}. 
}  (Further theoretical
motivations
for considering
such combinations are developed in section \ref{sec:forecasts}.) We introduce here
the shorthand notation $\hb_{0,1}$ to more compactly express the matrices involved with components $H^{(0,1)}_{ij}$.
The quantity $\Tr[\hb_{0}\hb_{0}]$ does not depend
on the velocity $\bv$ among frames,
and its magnitude depends only on the angle $\zeta$ between the stars, as defined   in eq \eqref{defoy}. (See 
also  Appendix \ref{Trace_Appendix} for analytical expressions  of
the combinations we plot.)  In the upper left panel of  Figure \ref{Maps_H}  we analyse the quantity $\Tr[\hb_{0}\hb_{0}]$
in the case 
where one of the star directions $\bf{n}$ points to  the centre of the sky map; the function reaches its local maxima at $\zeta=0^{\circ}$ and $\zeta\approx 105.6^{\circ}$, and has two roots at $\zeta=180^{\circ}$ and $\zeta\approx 57.10^{\circ}$. Such features
are also represented   on the left panel of Figure \ref{Plots_H}. Next, the quantity $H^{(1)}_{ij}$ controls the sensitivity to the kinematic dipole, and depends not only on the angle between the stars, but also on the angle between the velocity $\bv$ and the stars' directions. See eq \eqref{eq:H1ij}. 
In the lower panels of Figure \ref{Maps_H}
we present two maps to represent the properties
of this function. Complex patterns for the ORF
sensitivity to GW arise, depending
on the direction of star positions in the sky. Nevertheless, we can consider a simple scenario where $\bn$ is parallel to $\bv$. This assumption simplifies the expression for $\Tr[\hb_{1}\hb_{1}]$, 
and allows us to appreciate the system sensitivity to SGWB properties only  in terms of the angle separation $\zeta$ among the stars. In the upper right panel of  Figure \ref{Maps_H} the red band represents the regions of higher  sensitivity. The maximum of the function is reached at $\zeta \approx 119.45^{\circ}$, while the blue regions represent the lowest values of the function, including  its three roots at $\zeta=0, \pi$ and $\zeta \approx 73.14^{\circ}$. The resulting behaviour is also represented in   the right panel of Figure \ref{Plots_H}.

\subsection{Cross-correlated overlap
functions of
 astrometry and  PTA}\label{sec_xcorr}

In the future, precise
astronomical observations
will improve  astrometry measurements, making them
 more sensitive to GW.  The
   synergy of astrometry and  PTA
will be instrumental   for increasing the
sensitivity to the
properties of the SGWB. It
is therefore important
to analytically compute the overlap
functions associated with cross
correlations between astrometry
deflections and pulsar timings, also including effects of SGWB
anisotropies.
These topics have been explored
in \cite{Mihaylov:2018uqm,Qin:2018yhy,Inomata:2024kzr}. We present
here  covariant 
expressions for the overlap functions  of cross correlations, which are 
convenient for numerical
analysis and for the forecasts
we carry on in section \ref{sec:forecasts}. 

We consider two-point correlation
functions of deflections
of star positions, and of
pulsar time delays $z(t)\,=\,\Delta T(t)/T(t)$, where
$T(t)$ is the total time the radio signal needs for travelling
from the pulsar source to the detector. We denote  with $\bx$
the unit vector  pointing from
the Earth towards the specific pulsar being monitored. The time delay induced
by a GW propagating along
the direction $\bp$ results (see e.g. \cite{Maggiore:2018sht})
\be
z(t)\,=\,\frac12\,\frac{x^i \,x^j}{1+\bx \cdot \bp}\,h_{ij}(t,0)\,,
\ee
where we
include
the Earth
term only \cite{Maggiore:2018sht}. 
This piece of information, together
with the results of section \ref{sec:astrov}, allow
us to compute the correlation
functions between star deflection
$\delta n^i({\bf n},t)$ and $z(t)$,
induced by the presence
of a SGWB. It reads
\bea
\langle \delta n_i({\bf n},t)
z(t)\rangle
\,=\,\frac{3\,H_0^2}{64 
\pi^3}
\int df \frac{\bar \Omega_{\rm GW}(f)}{f^3}\,\left[
K_i^{(0)}({\bf n}, {\bf x})+\beta\,(4-n_\Omega)\,K_i^{(1)}({\bf n}, {\bf x}, {\bf v})
\right]\,.
\eea
In analogy with the formulas
developed
in section \ref{sec:astrov}, 
the quantities
 $K_i^{(0)}$ and $K_i^{(1)}$ 
denote respectively 
the cross-correlated
overlap reduction functions relative
to the SGWB monopole,
and kinematic dipole.
Using the same notation
of previous section,
they 
 are expressed in terms of the angular integrals
\bea
\label{defk0}
 K^{(0)}_{i} (\bn, {\bf x})
 &=& \int d^2 \Omega_{\bp}\,{\cal R}_{ikl} (\bn, \bp) \,
 P_{kl rs}\,\frac{x^r x^s}{1+ {\bf x}\cdot \bp}\,,
 \\
 K^{(1)}_{i} (\bn, {\bf x}, \bv)
 &=& \int d^2 \Omega_{\bp}\, ( \bp
 \cdot
 {\bf v})
  \,\,{\cal R}_{ikl} (\bn, \bp) \,
 P_{kl rs}\,\frac{x^r x^s}{1+ {\bf x}\cdot \bp}\,.
\eea
These integrals  can be computed straightforwardly through 
 contour integration
 in the complex
 plane. It is convenient to express
 the results in terms of the quantity (not to be confused with the analogous 
 quantity \eqref{defoy} of Section \ref{sec:astrov})
 \be
 y\,=\,\frac{1- \bn \cdot \bx}{2}\,,
 \ee
 related to the angle between
 the direction of the monitored star and  pulsar.
 The ORF of the
cross-correlation monopole reads
\bea
K^{(0)}_{i} (\bn, {\bf x})
 &=& \frac{16 \pi}{3}\frac{(1-2 y)\,n_i-x_i}{4 y(1-y)}
 \,\left( 2 y-2 y^2+3 y^2 \ln (y)\right)\,,
 \label{eq:K0i}
\eea
a formula equivalent
to the one found in  \cite{Mihaylov:2018uqm}, although
expressed in another form.

The overlap function for
the kinematic dipole is more
easily expressed in terms of 
  two vectors ${\bf A}_{1,2}$
orthogonal to the star direction $\bn$:
\bea
{\bf A_1}&=&{\bf n}\times {\bf x} 
\,,
\\
{\bf A_2}&=&  {\bf n}\times {\bf v} 
\,.
\eea
Notice the identity
$
{\bf A_1}\cdot
{\bf A_2}\,=\,\bx \cdot \bv-(\bn \cdot
\bv)(\bn\cdot \bx)
$. 
We parameterize the dipolar
overlap function as
\bea
K^{(1)}_{i} (\bn, {\bf x}, \bv)
 &=& 
 b_1 \,{ A_{1\,i}}+b_2 \, { A_{2\,i}}
 \label{eq:K1i}
 \,.
\eea
 
We denote ${\bf A_1}\cdot {\bf v}=(A_1 v)$,
etc. 
The two coefficients $b_{1,2}$ can 
be computed by contour integration, and result~\footnote{The  expression \eqref{eq:K1i} is
valid for all values 
of angles among the vectors involved, apart from when $\bn \cdot
\bv=0$. Then the 
vector ${\bf A}_2$ vanishes, as well
as the product ${\bf A}_1\cdot \bv$. The limit
 $\bn \cdot
\bv\to 0$ is hence delicate.
For this
specific case of velocity
vector $\bv $ parallel
to the star direction $\bn$, the expression
for the dipole response function
reduces to
\bea
\label{eq:soldk1a}
K^{(1)}_{i} (\bn, {\bf x})
 &=& \frac{2 \pi   ({n_i} (1-2 y)-{x_i})}{3 (1-{y})}
 \,\left(({y}-1)(6  {y}+1)-6 {y} \ln ({y})\right) 
 \label{K1i_specialC}
\,.
\eea}

\bea
b_1&=&\frac{\pi \left((A_1 v)^2 (1 - 12 y) + (A_1 A_2) \left((A_1 A_2) + 12 (A_1 A_2) y^2 + 4 (n v) y (1 + y (5 - 6 y)) \right)\right)}{6 (A_1 v) (1 - y) y}
\nonumber
\\
&&+ \frac{2 \pi \left((A_1 A_2)^2 - (A_1 v)^2 + 2 (A_1 A_2) (n v) (1 - y)\right) y \ln(y)}{(A_1 v) (1 - y)^2}\,,
\\
b_2&=&-\frac{2 \pi \left((A_1 A_2) + 12 (A_1 A_2) y^2 + 4 (n v) y \left(1 + y (5 - 6 y)\right)\right)}{3 (A_1 v)} 
\nonumber
\\
&&
- \frac{8 \pi \left((A_1 A_2) + 2 (n v) (1 - y)\right) y^2 \ln(y)}{(A_1 v) (1 - y)}\,.
\eea

\smallskip

Analogously to our discussion towards the end
of section \ref{sec:astrov},
 we can
visualise the 
ORF
for  astrometry working in synergy with PTA experiments. They are controlled
by the combinations 
 $\mathbf{K}_{0} \mathbf{K}_{0}^{T}$ and $\mathbf{K}_{1} \mathbf{K}_{1}^{T}$. We denote more compactly the vectors $K^{(0,1)}_{i}\equiv \mathbf{K}_{0,1} $.

\begin{figure}[h!]
    \centering                          \includegraphics[width=0.45 \linewidth]{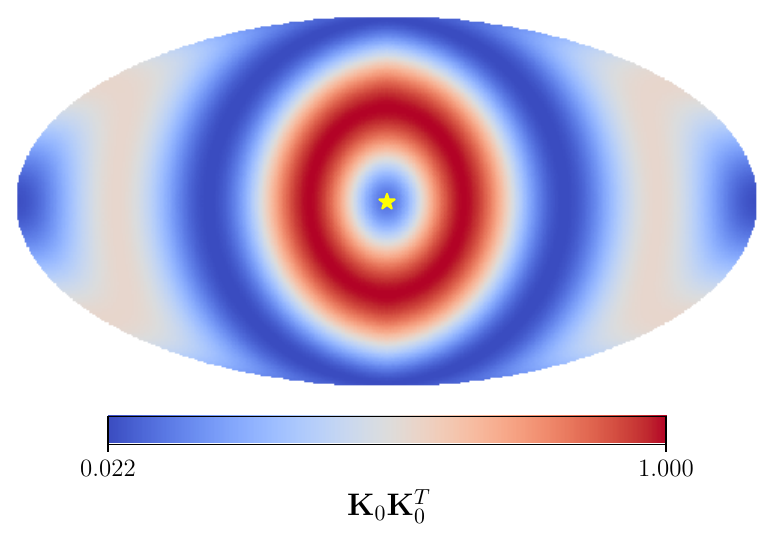}   \includegraphics[width=0.45 \linewidth]{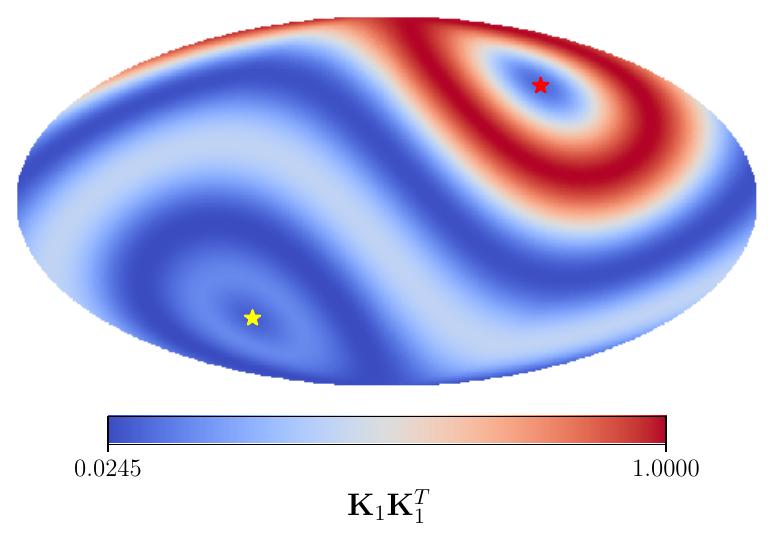}
    \includegraphics[width=0.45 \linewidth]{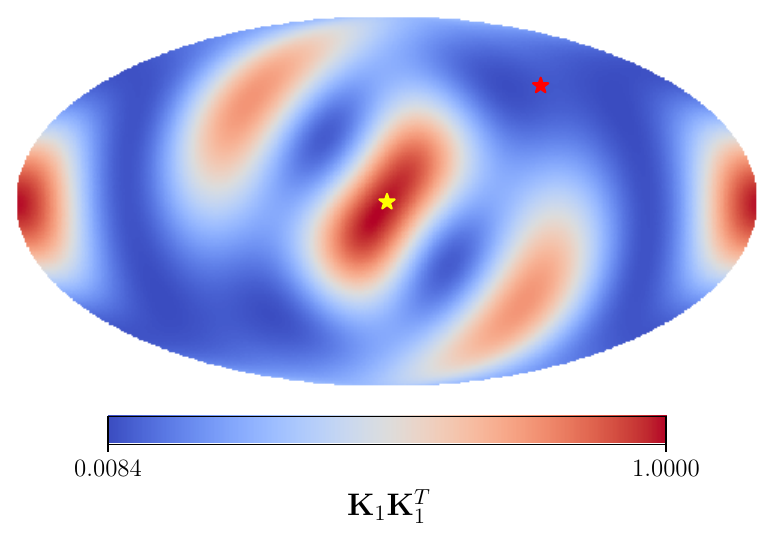}
    \includegraphics[width=0.45 \linewidth]{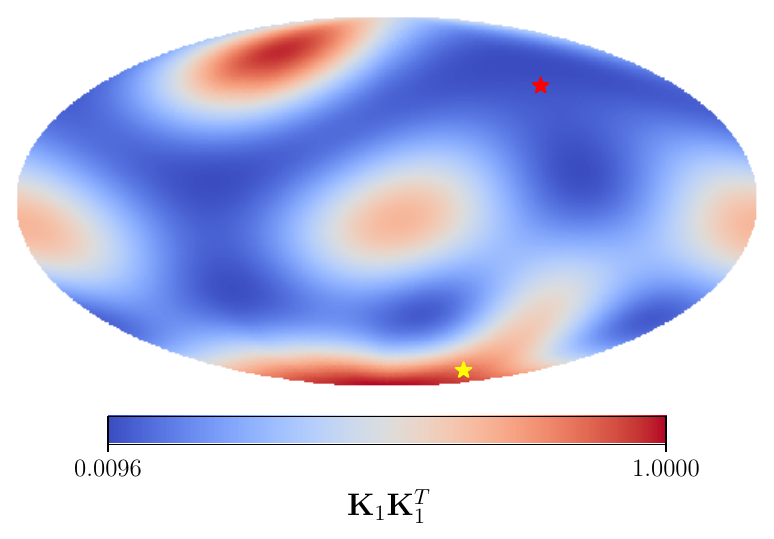}
    \caption{ \small The quantities $\mathbf{K}_{0} \mathbf{K}_{0}^{T}$, and
$\mathbf{K}_{1}\mathbf{K}_{1}^{T}$ response to stars and pulsars positions. The dipole direction $\bv$ (red star) is chosen in the direction  $(l,b)=(264^{\circ},48^{\circ})$ in galactic coordinates. Each panel shows a different choice of $\bn$, while the pulsars positions $\bx$ scan over  each pixel of the map.
  {\bf Upper left panel}: $\bn$ towards  $(l,b)=(0,0)$. 
 {\bf Upper right panel}: $\bn$ towards $-\bv$. {\bf Lower left panel}:  $\bn$ at $(l,b)=(0,0)$. {\bf Lower right panel}: $\bn$ towards  $(l,b)=(270.21^{\circ},-75.45^{\circ})$.}
    \label{Maps_K}
\end{figure}

\begin{figure}[h!]
    \centering            \includegraphics[width=0.45 \linewidth]{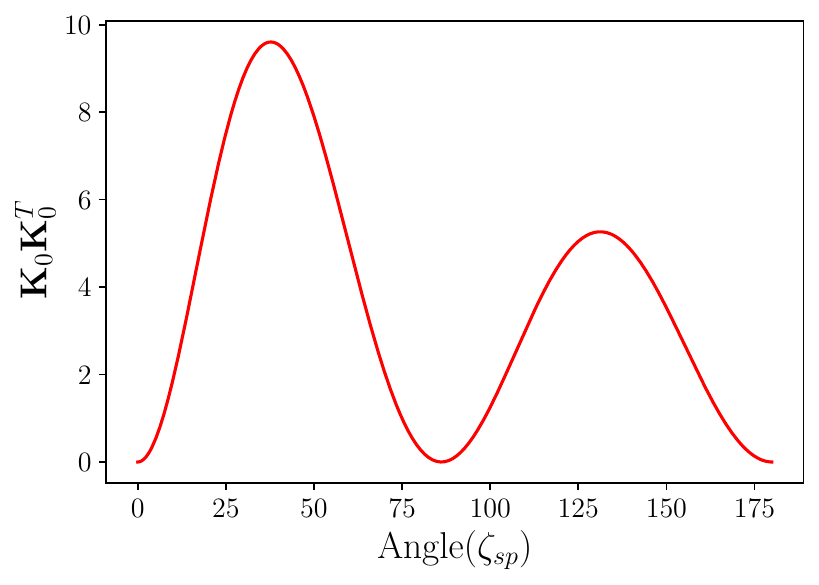}   \includegraphics[width=0.45 \linewidth]{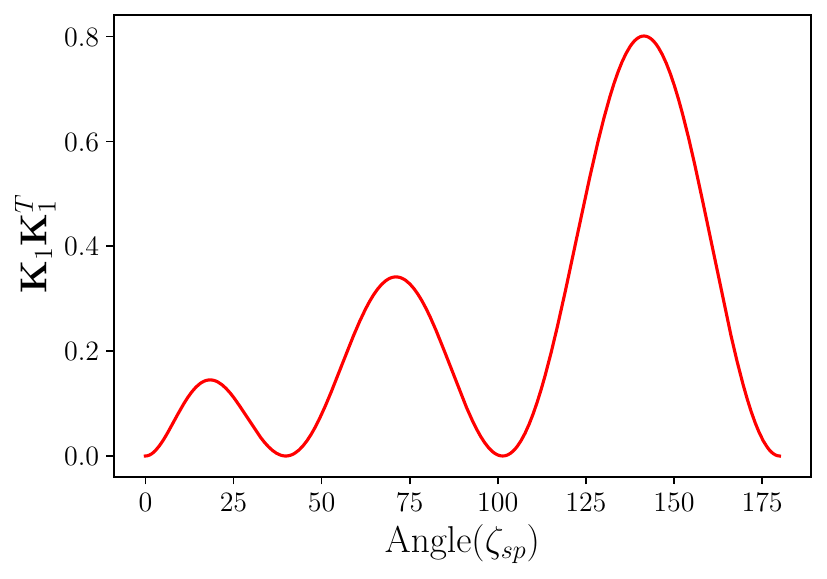}
    \caption{ \small Overlap functions response in terms of the angle between the stars and pulsars. {\bf Left:} Angular dependence for
$\mathbf{K}_{0}  \mathbf{K}_{0}^{T}$. {\bf Right:} The quantities   $\mathbf{K}_{1}  \mathbf{K}_{1}^{T}$ when choosing $\bn = - \bv$. }
    \label{Plots_K}
\end{figure}

We call $\zeta_{sp}$ the angle
between star and pulsar directions, $\cos \zeta_{sp}={\bn}\cdot \bx$.
We represent
 a simple example  in the upper left panel of Figure \ref{Maps_K} with $\bf{n}$ pointing towards $(l,b) = (0,0)$; the roots of the function $\mathbf{K}_{0} \mathbf{K}_{0}^{T}$ are at $\zeta_{sp}=0, \pi$ and $\zeta_{sp} \approx 86.14^{\circ}$, while its local maxima are located at $\zeta_{sp} \approx 37.13^{\circ}$ and $\zeta_{sp} \approx 132.195^{\circ}$,  and are represented in the left panel of Figure \ref{Plots_K}. The lower panel of Figure \ref{Maps_K} shows two scenarios for the general expression for the function  $\mathbf{K}_{1} \mathbf{K}_{1}^{T}$ (see also Appendix \ref{Trace_Appendix}), which depends on the angle 
  between star and pulsar  and the angle between the velocity and pulsars given by $vx = \bv \cdot \bx$; on the left we choose $\bf{n}$ at $(l,b) = (0,0)$, and on the right, $\bf{n}$ is at  position $(l,b)=(270.21^{\circ},-75.45^{\circ})$ (a direction 
   chosen randomly). The sensitivity patterns  are  complex. However, in the specific case of the star direction $\bf{n}$ pointing towards the
 direction $-\bv$ we obtain a simple formula, depending  only on the angle  between star and pulsar. This last case is shown on the right upper panel of Figure \ref{Maps_K}, where the bluest, less sensitive regions showing  the positions of objects at relative   angle $\zeta_{sp} = 0, \pi$ and $ 39.82^{\circ}, 101.37^{\circ}$, corresponding
 to minimal sensitivity to
 dipole. The  red region shows the positions of the pulsars where the function 
 $\mathbf{K}_{1} \mathbf{K}_{1}^{T}$ reaches its maximum at $\zeta_{sp} \approx 142.29^{\circ}$. There are also local maxima at $\zeta_{sp} \approx 17.97^{\circ}, 71.88^{\circ}$, which are more apparent  on the right panel  of Figure \ref{Plots_K}.  Notice
 that the quantities ${\bf K}_{0,1}$ corresponding
 to the synergetic
 ORF vanish in the concident
 limit of the star aligned
with the pulsar direction. One may understand this as follows -- the star deflection always lies in the plane perpendicular to the line of sight. On the contrary, the pulsar timing residual arises from the change in length along the line of sight. Thus for pulsar and star in same direction the two effects are  perpendicular to each other.


\newpage
\section{Fisher forecasts}
\label{sec:forecasts}

Armed with the covariant, 
analytical expressions we derived  
for the ORF of astrometry and PTA systems -- including the effects of
kinematic
anisotropies -- in this
section we investigate
the prospects
of future
experiments to characterize
the SGWB. We are interested
in measuring the SGWB amplitude,
its spectral tilt, as well
as the magnitude of
a possible dipolar anisotropy characterizing
the SGWB.
 We quantify how
the results depend on the number of astronomical
objects we measure, as well as on the precision of their  measurement.

\smallskip


{In this section we use the Fisher matrix approach to forecast the sensitivity of astrometry, individually and jointly with PTA, to the SGWB properties such as the amplitude, spectral index and dipole anisotropy.}
In appropriate limits and special cases, the Fisher
matrix obeys interesting scaling relations which make our analysis particularly simple. In general though, numerical work is needed
to evaluate  at what extent the synergy between astrometry and PTA
improve the sensitivity to
SGWB properties. We carry out
such analysis in sections \ref{sec:Fisher_astro_only}, \ref{sec:PTA_AstrometryForecast} and \ref{sec_syndip}, making use of the weak signal approximation for astrometry to simplify the numerical calculations for a large number of stars. 

\subsection{Astrometry only}
\label{sec:Fisher_astro_only}
We start by discussing
the case of astrometry only.
We expand the deflection correlation
of eq~\eqref{eq_secexp} in a basis with coefficients $p_n$,\footnote{In principle, these basis coefficients can also be spherical harmonics coefficients or individual pixels used to discretize the SGWB intensity map. Our analysis represents a special case where we only expand the SGWB intensity up to the dipole term, assuming a known dipole direction.}
\begin{align}
\label{est2ddef}
\qev{\delta n_a^i \, \delta n_b^j} \equiv C^{ij}_{ab} = \sum_{n=0,1} p_n H_{ab,n}^{ij} + N_{ab}^{ij}
\end{align}
The indexes $a, b$ identify the two stars. $i, j$ the
three-dimensional vector
components, and the index $n$ runs over $0,1$  indicating 
monopolar and dipolar contributions to the signal. 
We include the effects of noise {through} the noise matrix
$ N_{ab}^{ij}$. The quantities
$p_n$ and $H_{ab,n}$ are controlled
by the properties of the SGWB, as well as the astrometry ORF. See equation \eqref{eq_secexp}
and the discussion that follows for the definitions. {The angle brackets on the LHS denote the expectation value of the correlation of the $i$-th deflection of star $a$ with the $j$-th deflection of star $b$. 
To keep our notation compact we now use bold symbols to represent the signal covariance as well as the noise matrix, i.e.
\begin{align}
\label{est2ddef_2}
C^{ij}_{ab} \equiv \cb = \sum_{n=0,1} p_n H_{ab,n}^{ij} + N_{ab}^{ij} \equiv \sum_{n=0,1} p_n \hb_n + \nb\,. 
\end{align}
Thus, in the following the trace of the product of two bold quantities represents a sum over the $i,j$ and $a,b$ indices, i.e.
\begin{align}
    \Tr[\boldsymbol{A}\boldsymbol{B}] = \sum_{ij,ab}A^{ij}_{ab}B_{ab}^{ij}\,.
\end{align}
}
At this stage,
we do not need to specify whether we work either in time or in frequency domain. 
\smallskip

{Our aim is to determine
the how well we can measure the quantities $p_{0,1}$, and
extract from this information
the properties of the SGWB: its monopole amplitude
$\Omega_{\rm GW}$ 
as well as the value of the parameter
$\beta$ controlling kinematic anisotropies, using a Fisher matrix approach.\footnote{{We also discuss the optimal estimators for the SGWB monopole and dipole in \cref{sec:Astrometry_estimator}}.} However, before proceeding to build the Fisher matrix for the monopole and dipole amplitudes, }
we  note that the matrix $C$ can not be directly be used as a covariance matrix,  since it is singular. This stems from the fact that the three Cartesian deflection components are not independent
since
 observations
 get projected on to the two-dimensional celestial
 sphere. Hence,
the star deflections can be entirely described in terms of  two angles $\delta\theta$ and $\delta\phi$~\cite{Mihaylov:2018uqm}. 
Let $P$ be the matrix that converts the deflections from 3D Cartesian to 2D polar coordinates, i.e.
\begin{align}
    \begin{pmatrix}
        \delta\theta \\
        \delta\phi 
    \end{pmatrix} = P \cdot
    \begin{pmatrix}
        \delta x \\
        \delta y \\
        \delta z
    \end{pmatrix}\,.
\end{align}
The matrix $P$ is  used to convert the original correlation matrix for the Cartesian deflections to a correlation matrix in terms of angular deflections.
For a single star, it is given by~\cite{Mihaylov:2018uqm}
\begin{align}
    P = \begin{bmatrix}
        0 & 0 & \frac{1}{\sqrt{1-z^2}} \\
        -\frac{y}{x^2 + y^2} & \frac{x}{x^2+y^2} & 0
    \end{bmatrix}\,.
\end{align}
For the full system of $N$ stars, the projection matrix $R$ can be written in block diagonal form~\cite{Mihaylov:2018uqm}
\begin{align}
   R =  \begin{bmatrix}
        P_1 & 0 &  \ldots & 0\\
        0 & P_2 & \ldots & 0\\
        \vdots & \vdots & \ddots & \vdots \\
        0& 0& \ldots &  P_n
    \end{bmatrix}\,.
\end{align}
This can be used to obtain the angular deflection covariance matrix from the Cartesian one
\begin{align}
    C^{\delta \vec{\theta}}_{ij} = R\cdot C_{ij}^{\delta \vec{x}} \cdot R^T\,,
\end{align}
which is the actual covariance matrix that will be used for calculations to follow.
{For zero mean, Gaussian distributed astrometric deflections, the likelihood $\lc$ takes the form 
\begin{align}
    \label{eq:astro_likelihood}
    -2\ln \lc = \delta n^i_a [C^{-1}]^{ij}_{ab}\delta n^j_b + \ln \det \cb\, + n\ln 2\pi.
\end{align}
How well we can determine the parameters that describe our signal is controlled by
the Fisher matrix, defined as~\cite{Tegmark:1996bz}
\begin{align}
\label{deffim}
        \fc_{ij} = \left\langle -\frac{\partial \ln \lc}{\partial \theta_i \partial \theta_j} \right \rangle\,,
    \end{align}
    given in terms of the expectation value of second derivatives of $ \ln {\cal L}$. The Fisher matrix sets a lower bound on the precision with which these parameters can be measured, \mbox{$\Delta \theta_{i,\rm min} = \sqrt{[\fc^{-1}]_{ii}}$}.
For the likelihood in \eqref{eq:astro_likelihood}, the Fisher matrix reads~\cite{Tegmark:1996bz} 
\begin{align}
    \fc_{ij} &= \frac{1}{2}\Tr\left[C^{-1}C_{,i}C^{-1}C_{,j}\right]\,
\end{align}
where $C_{,i}$ denotes the derivative of $C$ w.r.t the parameter $\theta_i$. Using the definition of the covariance \eqref{est2ddef}, we see that 
\begin{align}
    C_{,i} =\sum_n \frac{\partial p_n }{\partial \theta_i}\hb_n \,.
\end{align}
Specialising to the case where our signal parameters $\theta_i$ correspond to the monopole and dipole amplitude $p_0$
and $p_1$ as in eq \eqref{est2ddef}, the Fisher matrix reads
\begin{align}
      \fc_{ij} = \frac{1}{2}\Tr \left[ C^{-1} H_i C^{-1} H_j \right]\,,
    \label{eq:Fisher_astro_no_approx}
\end{align}
where $i,j$ run from 0 to 1.} In the noise dominated limit with a diagonal noise covariance matrix \mbox{$N_{ab}^{ij} = \sigma^2_N \delta_{ab}\delta^{ij}$}, we obtain the Fisher matrix $\fc$ for $p_0, p_1$ as
\renewcommand{\arraystretch}{1.5}
\begin{align}
\label{eq:astro_fisher}
    \fc = \frac{1}{2 \sigma^4_N}\begin{pmatrix}
       \Tr[\hb_{0}\hb_{0}]\quad  &  \Tr[\hb_{0}\hb_{1}] \\
       \Tr[\hb_{0}\hb_{1}]\quad & \Tr[\hb_{1}\hb_{1}]
    \end{pmatrix}\,.
\end{align}
\begin{figure}
\label{fig:Tr_HH}
    \centering
    \includegraphics[width=0.47\linewidth]{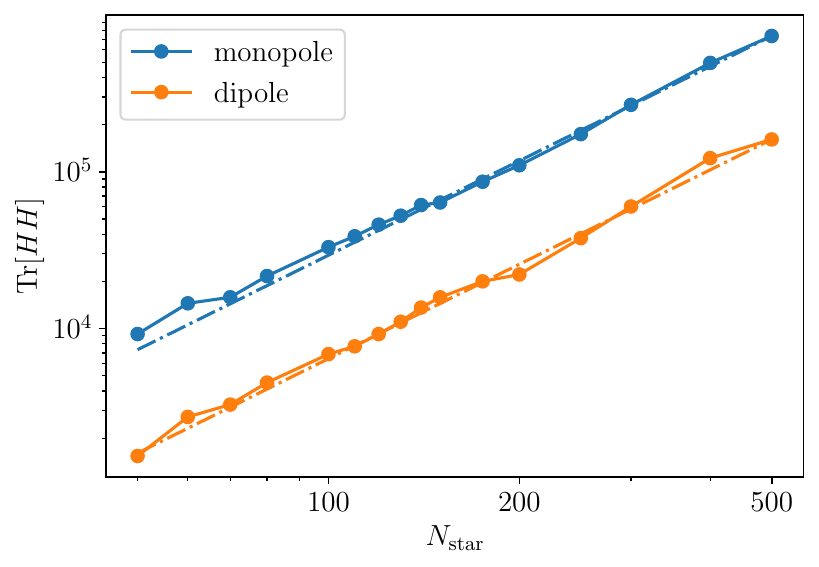}
    \caption{\small The components of the Fisher matrix \eqref{eq:astro_fisher} are evaluated numerically as a function of the number of stars with the stars uniformly distributed across the sky. The dashed lines have slope $N_{\rm star}^2$ and pass through the numerically evaluated result for $N_{\rm star}=500$. This plot numerically confirms eqs  
    \eqref{eq:traceH} and \eqref{eq:traceHa}.
    }
    \label{fig:Tr_scaling}
\end{figure}
Hence, in the noise-dominated regime, the Fisher matrix  depends only on the ORF of the monopole and dipole of the SGWB,   discussed 
 in Section \ref{sec:theory_ORF}.

\begin{figure}[h]
    \centering    \includegraphics[width=0.47\linewidth]{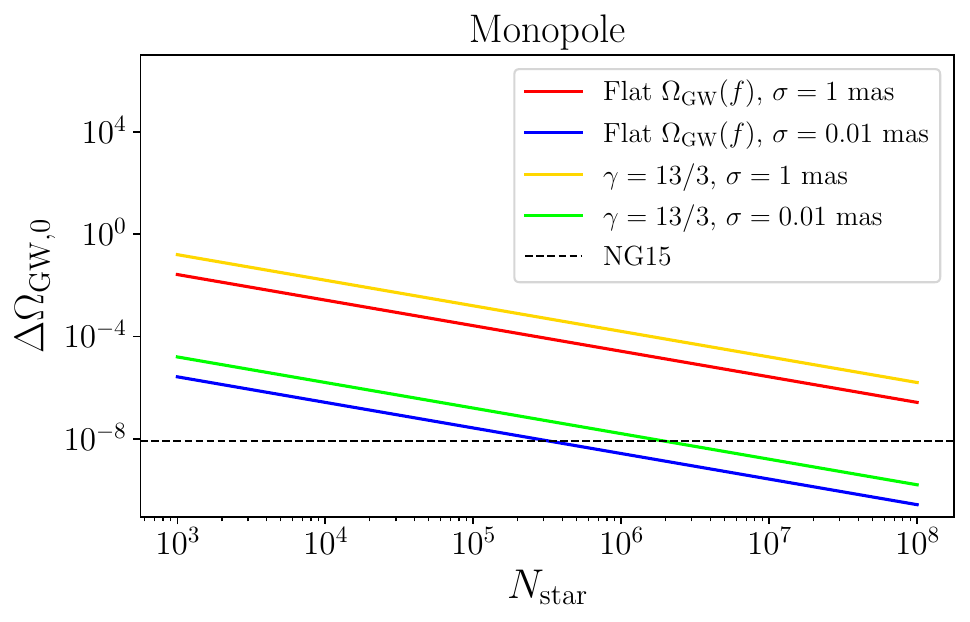}
    \includegraphics[width=0.47\linewidth]{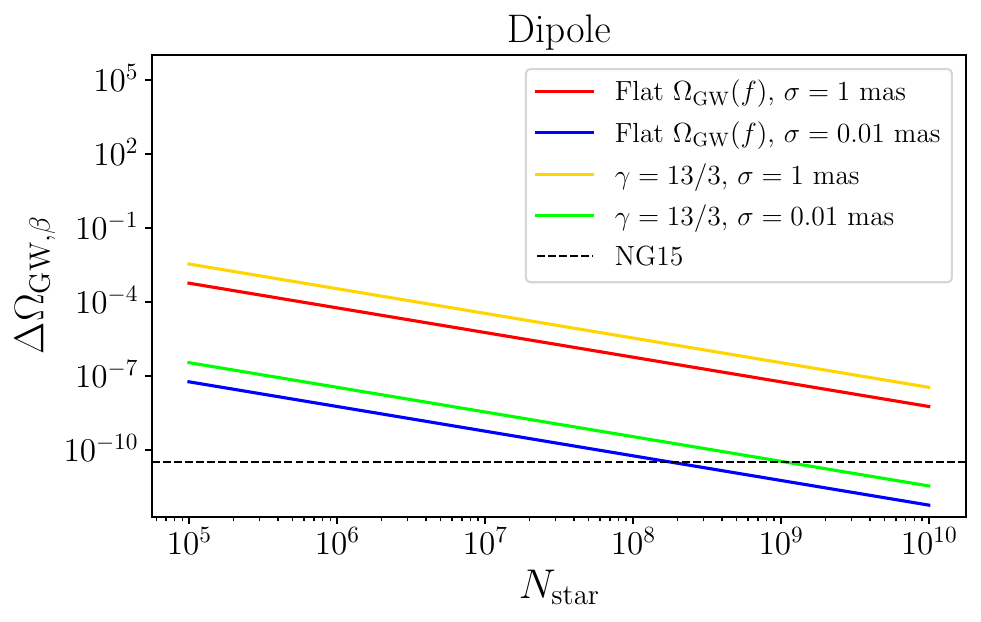}
    \caption{\small Forecasts for the
    magnitude of the SGWB energy density and dipolar anisotropy, as measured with astrometry. See the discussion after eq \eqref{eq:traceHa}. {\bf Left}: The error $ \Delta \Omega_{\rm GW, 0}$ associated with the monopole estimator $p_0$. {\bf Right}: $\Omega_{\GW, \beta}$ associated with the dipole estimator $p_1$. We define $\Omega_{\GW, \beta} =  \beta (4-n_{\Omega}) \Omega_{\rm GW ,0}$. {The dashed lines correspond to $\Delta\Omega_{\rm GW, 0}\approx 8.05 \times 10^{-9}$, which is the value of the energy density with the NG15 results ~\cite{NANOGrav:2023ctt}, and $\Delta\Omega_{\GW, \beta}\approx 3.3 \times 10^{-11}$ . 
    }
    }
\label{fig:Forecast_Omega}
\end{figure}

The entries of the
Fisher matrix \eqref{eq:astro_fisher} simplify under certain
hypothesis. For example, we find the
following scaling  relations which hold in the limit of large $N_\mathrm{star}\gg 100$ stars distributed uniformly across the sky
(see
Figure \ref{fig:Tr_scaling}):
\begin{align}
    \label{eq:traceH}
    \fc_0 &\equiv \Tr[\hb_{0}\hb_{0}] \simeq 3\times N_\mathrm{star}^2 \,,\\
    \fc_1 &\equiv \Tr[\hb_{1}\hb_{1}] \simeq 0.65\times N_\mathrm{star}^2 \,.
    \label{eq:traceHa}
\end{align}
In the same limit, 
  the cross terms 
depending on {$\Tr[\hb_{0}\hb_{1}]$}
vanish in eq \eqref{eq:astro_fisher}.  
In practice, to perform   Fisher forecasts by means
of Eq \eqref{eq:astro_fisher} 
we focus on the frequency domain, performing a frequency binning with $\Delta f = 1/T_{\rm obs},$ with $ T_{\rm obs}=15\;\rm years$, and summing over the Fisher matrices at each frequency bin.
We relate the value of the GW energy density $\Omega_{\rm GW}$ to the intensity through equation 
\eqref{rel_ogwi}. We assume for the latter a power-law ansatz $I(f) = I_0 (f/f_*)^{2-\gamma}$, and report our results
in Fig \ref{fig:Forecast_Omega} for different values of spectral slope, also indicating the detection
threshold indicated by the IPTA joint analysis~\cite{InternationalPulsarTimingArray:2023mzf}. 

The noise in the frequency domain is given by $\sigma^2_N (f) =  2\sigma^2 T_{\rm cad} $, {where we have assumed a white noise component only}. We take different values of $\sigma$ as shown in  Figure \ref{fig:Forecast_Omega} and the observational cadence $T_{\rm cad} = \mathrm{year}/15$,  corresponding approximately to the observational cadence of Gaia~\cite{Gaia:2016zol}, 
 {i.e. $T_{\rm cad}$ represents the time interval (in seconds) between successive observations of a particular star.}

{Assuming noise dominated covariance, the uncertainties for the intensity monopole amplitude $I_0$ and the dipole amplitude $I_{\beta}\equiv(1-n_I)\beta I_0$ \footnote{{ Note the relation $n_{\Omega} = n_{I} + 3$ arising from \eqref{rel_ogwi}.}}, take the following form
\begin{align}
        \fc_{I_0 I_0} &= \frac{T_{\rm obs} }{2\sigma^4_N }\frac{\,\Tr[\hb_{0}\hb_{0}]}{(4\pi)^2}\sum_{f} (f/f_*)^{2n_I} \Delta f\,, \\
        \fc_{I_{\beta} I_{\beta} } &= \frac{T_{\rm obs} }{2\sigma^4_N }\frac{\,\Tr[\hb_{1}\hb_{1}]}{(4\pi)^2}\sum_{f} (f/f_*)^{2n_I} \Delta f\,,
\end{align}
with $n_I = 2 - \gamma$. 
To obtain a scaling relation, we replace the discrete sum by an integral,
\begin{align}
   \sum_{f} (f/f_*)^{2n_I} \Delta f \approx \int_{f=1/T_{\rm obs}}^{1/T_{\rm cad}} (f/f_*)^{2n_I} df\,.
\end{align}
Given that $n_I = -7/3 <0$ \footnote{{The fiducial value $n_{I} = -7/3$ adopted here corresponds to the theoretical expectation for supermassive black hole binaries in circular orbits~\cite{Phinney:2001di}.} }, the integral is dominated by the lower frequencies and can be approximated (for $n_I\neq -1/2)$ by
\begin{align}
    \left[\int_{f=1/T_{\rm obs}}^{1/T_{\rm cad}} (f/f_*)^{2n_I} df \right] \approx \left[-\frac{f_{\rm min}}{2n_I+1} \left(\frac{f_{\rm min}}{f_*}\right)^{2n_I}\right]\,.
\end{align}
Since, for a given parameter $\theta$, $\Delta \theta = \sqrt{[\fc^{-1}]_{\theta \theta}}$, we obtain
\begin{align}
    \Delta I_0 &= \sqrt{\frac{2}{3 |2n_I+1|}} \frac{4\pi\sigma^2_N}{N_{\rm star}} \left(\frac{T_{\rm obs}}{1 \,\mathrm{year}}\right)^{n_I} \,, \\
    \Delta I_{\beta} &= \sqrt{\frac{2}{0.65 |2n_I+1|}} \frac{4\pi\sigma^2_N}{N_{\rm star}} \left(\frac{T_{\rm obs}}{1 \,\mathrm{year}}\right)^{n_I}  \,,
\end{align}
where we used the fact that $f_{\rm min} = 1/T_{\rm obs}$ and $f_* = 1 / \mathrm{year}$. Using the relation between the intensity and energy density \eqref{rel_ogwi}, allows us to convert these expressions to uncertainties in $\Omegagw$. We find
\begin{align}
    \Delta \Omega_{\rm GW,0(\beta)} &= 1.6 (3.4) \times 10^{-8}\sqrt{\frac{|2n_I+1|}{4/3}} \left(\frac{\sigma}{0.01}\right)^2 \left(\frac{T_{\rm cad}}{\mathrm{year}/15}\right) \left(\frac{10^6}{N_{\rm star}}\right) \left(\frac{T_{\rm obs}}{15 \,\mathrm{year}}\right)^{n_I}\,,  
\end{align}
We emphasise that these relations hold only in the noise dominated (weak-signal) limit and for sufficiently red tilted SGWB and assuming that this tilt is known and not itself being inferred from the data.
}

Under the assumptions leading
to eqs \eqref{eq:traceH} and \eqref{eq:traceHa} we find
that  astrometric surveys with $0.01$ milli-arcsecond (mas) precision which monitor $N>10^6$ stars may be competitive with PTA
experiments, in terms of sensitivity to the SGWB. The detection of the kinematic dipole, whose amplitude is suppressed by a factor $\beta$ relative to the monopole, will accordingly require about $10^3$ times more stars. 
We should note, though, that for the green and blue lines shown in \cref{fig:Forecast_Omega}, the weak signal approximation likely breaks down, as we go towards a higher and higher number of stars ($N_{\rm star} \gg 10^6$). The noise level $\sigma = 0.01 $ mas roughly corresponds to the expected astrometric accuracy that will be achieved by Gaia DR5 for the brightest objects in the survey.\footnote{\url{https://www.cosmos.esa.int/web/gaia/science-performance}} Thus, the forecast sensitivity with this value of the noise is very unlikely to be achieved in real Gaia DR5 data but may be possible with future missions like Theia~\cite{Theia:2017xtk,Garcia-Bellido:2021zgu,Malbet:2022lll}.

\subsection{Forecasts:  astrometry in synergy with PTA}
\label{sec:PTA_AstrometryForecast}

We now turn our attention to estimators and forecasts for astrometry in synergy with PTA. In this section
we consider 
only the isotropic part of the background, {focusing on the amplitude and spectral tilt of the SGWB. The effect of the kinematic dipole is investigated in section \ref{sec_syndip}.  \\

Let $\delta t_a,\, a=1,2,\ldots N$, and $\delta \vec{\theta}_b,\, b=1,2,\ldots M/2$ be the Fourier transforms of timing residuals and angular deflections of $N$ pulsars and $M/2$ stars respectively.\footnote{The factor of two arises for the stars because they are each characterized by two angular deflections in the sky, while pulsars only by a single time delay.}  We work under the assumption of a Gaussian distributed  SGWB with zero mean. The joint covariance matrix at a given frequency can be written in block form as
\begin{align}
 C_{(N+M)\times (N+M)} =    \begin{bmatrix}
        A_{N\times N} & B_{N\times M} \\
        B^T_{M\times N} & D_{M\times M}
    \end{bmatrix}\,,
    \label{mat_corra}
\end{align}
and the joint likelihood as
\begin{align}
    -2 \ln L = (\vec{\delta t},\vec{\delta \theta})\cdot C^{-1}\cdot  (\vec{\delta t},\vec{\delta \theta})^T + \ln \det C + \frac{(M+N)}{2}\ln 2\pi\,.
\end{align}
As explained previously, we are going to split in frequency bins the total frequency interval we analyse. 

The individual sub-matrices $A,B,D$ denote the pulsar-pulsar, pulsar-star and star-star covariance matrices respectively. Notice
that they have very different dimensionalities, since we expect to monitor many more stars than pulsars (more on this later). 
They read 
\bea
\label{eq:auto_cross_block_def}
    A_{pq} &=& \frac{\gamma_{pq}I_f}{(4\pi f)^2} + \sigma^2_p \delta_{pq}\,, 
    \\
    B_{pa} & =& \frac{K_{p,a}I_f}{{(4\pi)^2} f}\,,
    \\
    \label{def_matD}
    D_{ab} &=& \frac{H_{ab}}{{4\pi}}I_f + \delta_{ab}\sigma^2_{a}\,.
\eea
The tensor $\gamma_{pq}$ corresponds the Hellings-Downs (HD) inter-pulsar correlation between pulsar time delays\footnote{{For the HD correlations, we adopt  the normalization convention of \cite{Ali-Haimoud:2020ozu} where $\gamma_{pp}=4/3$.}}, while $K_{p,a}$ and $H_{ab}$ denote the pulsar-star and star-star correlations studied in section \ref{sec:theory_ORF}. In general, the covariance matrices depend
explicitly on the SGWB intensity $I_f$ evaluated at the frequency bin under examination. (We use the same notation as \cite{Cruz:2024svc,Cruz:2024esk}.) 

The Fisher matrix for the joint forecast is given by the usual formula~\cite{heavens2009statistical}
\begin{align}
    \fc_{\alpha \beta} = \frac{1}{2}\Tr[C^{-1}C_{\alpha}C^{-1}C_{\beta}]\,.
\end{align}
where $C_{\alpha}$ denotes the derivative with respect to the parameter $\alpha$. We wish to investigate how the addition of the astrometric datasets can help to improve upon on the constraints on the SGWB amplitude  and spectral tilt, with respect to an analysis based only on PTA data. 

We first calculate the inverse of our covariance matrix. When the matrices $A$ and $D$ are both invertible, as is the  present case, the inverse is given by~\cite{lu2002inverses,horn2012matrix} 

\begin{align}
\label{eq:Cinv_block}
    C^{-1} = \begin{bmatrix}
        (A-BD^{-1}B^T)^{-1} & 0 \\
        0 & (D-B^T A^{-1}B)^{-1}
    \end{bmatrix} 
    \begin{bmatrix}
        \idn & -BD^{-1} \\
        -B^T A^{-1} & \idm
    \end{bmatrix}\,.
\end{align}
While the derivative $C_{\alpha}$ reads
\begin{align}
    C_{\alpha} = \begin{bmatrix}
        A_\alpha  & B_\alpha \\
        B^T_\alpha & D_\alpha
    \end{bmatrix}\,.
\end{align}

\subsubsection{Some concrete examples}
\label{sec_conex}

{To demonstrate  in concrete how the cross-correlation of astrometric and PTA data can be useful to characterize the SGWB signal -- even when the astrometric data alone are not very constraining -- we start with a simple {toy} example.  \\
\noindent {\bf Case 1}:
We parameterize the SGWB intensity as
\begin{equation}
I(f) = \frac{A_{\rm GW}^2}{2 {f_{\rm ref}}}\,\left(\frac{f}{f_{\rm ref}}\right)^{2-\gamma}\,,
\end{equation}
{using the quantities $A_{\rm GW}$ and $\gamma$ to make comparison with PTA analyses simpler,}\footnote{{We follow the conventions outlined in \cite{Cruz:2024svc} to relate these quantities, where $h_{c}(f)\equiv \sqrt{2fI(f)} = A_{\rm GW}(f/f_{\rm ref})^{\alpha}$, with $\gamma \equiv 3-2\alpha$.}} 
with $f_{\rm ref} = 1/\rm year$.
We choose PTA parameters so as to recover the level of constraints set by the IPTA joint analysis~\cite{InternationalPulsarTimingArray:2023mzf} with 15 years of observation time, roughly corresponding to $\log_{10} A_{\rm GW} = -14.6\pm 0.16,\, \gamma = 13/3\pm 0.45
$ at $95\%$ C.L. 
For the astrometric analysis, we take 1000 stars with identical noise level $\sigma^2_S = 2 [\Delta\theta_{\rm rms}]^2 T_{\rm cad}$ with $\Delta\theta_{\rm rms}=0.0002\,\rm mas$ and $T_{\rm cad}=\mathrm{year}/52$, i.e. with each star observed once per week.

Although {these numbers are not representative of typical astrometric datasets}, which have a larger number of stars, and at the same time larger noise levels, this {toy} example suffices to make our point {and is computationally feasible without making any approximations for the Fisher matrix}. {Both the pulsar and the star
configuration is supposed to be distributed uniformly across the sky. } 

\smallskip

\begin{figure}
    \centering
    \includegraphics[width=0.46\linewidth]{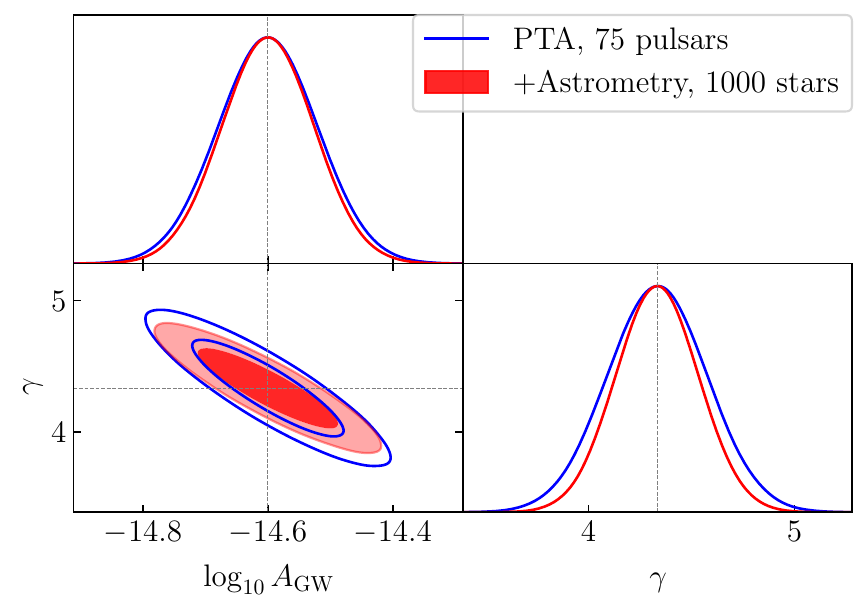}
    \includegraphics[width=0.45\linewidth]{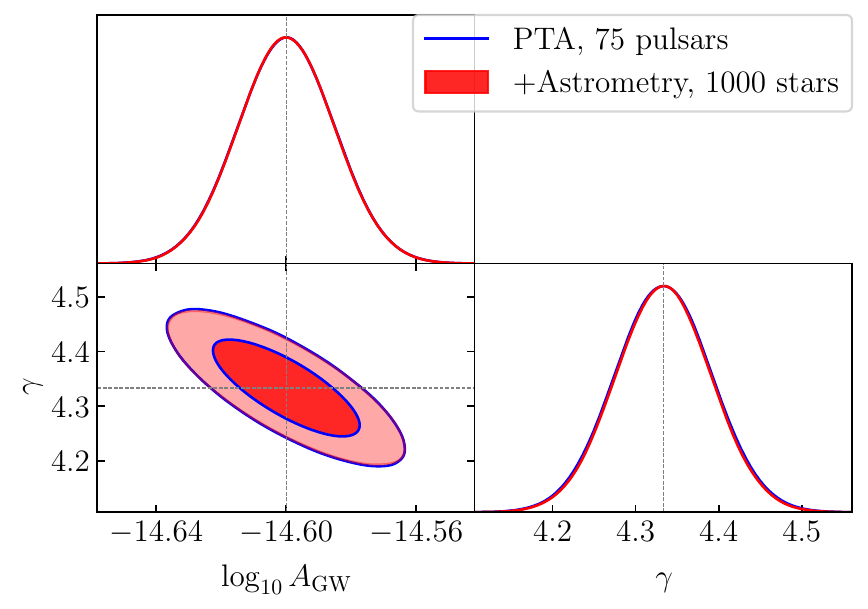}
     \caption{\small 
     Fisher forecast  for 
    the { Case 1} ({\bf left}) and Case 2 ({\bf right}) examples of section \ref{sec_conex}.}
    \label{fig:fisher_cross_toy1}
\end{figure}

{In the left panel of \cref{fig:fisher_cross_toy1}}, we  plot the Fisher forecast for the parameters $\log A_{\rm GW}$
and $\gamma$ within  this setup.
We learn  that the joint PTA+Astrometric dataset does provide an improvement over the PTA-only constraints, even when the astrometric data alone is not too informative. 

\noindent \textbf{Case 2:}
{We proceed by presenting one more example, plotted in the right panel of \cref{fig:fisher_cross_toy1}, keeping the same noise properties for the astrometry set, but  with the PTA system in the strong-signal (noiseless) regime. We see that the improvement in constraints with the addition of astrometry can be sizeable if the PTA datasets are not in the strong signal regime; however, it is much smaller if that is indeed the case.}

\subsubsection{A method for handling  large Fisher matrices}
\label{sec_methhand}

In the 
examples discussed above we are able to perform  the necessary computations  involving Fisher matrices and their inversion,
 given the relatively small size of astrometric covariance matrix involved. However, for realistic datasets which may contain over $10^6$ stars, inverting matrices 
of such dimensionality  
 can become demanding. To handle forecasts for such datasets, we develop  a simple method to perform the Fisher forecasts in weak astrometric signal limit without needing to resort to the full numerical calculations. The weak signal limit is justified since current astrometric datasets are very much in the noise dominated regime and this is likely to be the case in the future as well.\footnote{To be precise this requires $M I(f) \ll {4\pi} \sigma^2 (f)$ where $\sigma^2$ denotes the frequency domain noise power spectral density. This condition ensures that the effect of the off-diagonal terms in the covariance is suppressed compared to the diagonal terms.} Thus, we can approximate the matrix $D$ of equation \eqref{def_matD} as
\begin{align}
    D_{ij} \approx \delta_{ij}\sigma^2_i,\quad  [D^{-1}]_{ij} = \delta_{ij}\sigma^{-2}_i\,. 
\end{align}
We further assume all stars to have the same measurement noise $\sigma^2_i = \sigma^2_S$, as well as identical noise for each pulsar $\sigma^2_P$. 

} 

In the noise dominated approximation for the astrometric covariance, we write 

\begin{align}
        C^{-1} = \begin{bmatrix}
        (A-B\sigma^{-2}_S \idm B^T)^{-1} & 0 \\
        0 & (\sigma^{2}_S \idm - B^T A^{-1}B)^{-1}
    \end{bmatrix} 
    \begin{bmatrix}
        \idn & -B\sigma^{-2}_S \idm \\
        -B^T A^{-1} & \idm
    \end{bmatrix}\,.
\end{align}

 Note that we do not assume a weak signal limit for the pulsar covariance: in fact, we work with the full PTA covariance matrix. In fact, we focus on the relevant regime for PTA experiments, since current measurements already suggest PTA measurements to lie in the intermediate signal regime -- especially for the lower end of the  frequency range. To evaluate the trace we compute

\begin{align}
    C_{\alpha}C^{-1}  & = \begin{bmatrix}
        A_\alpha Q_1 & \;\;\;\; B_\alpha Q_2 \\
        B^T_\alpha Q_1  & \;\;\;\; D_\alpha Q_2
    \end{bmatrix}   
    \begin{bmatrix}
        \idn & -B\sigma^{-2}_S \idm \\
        -B^T A^{-1} & \idm
    \end{bmatrix}\,
    \\
    &=\begin{bmatrix}
         A_\alpha Q_1 -B_\alpha
         Q_2 \,B^T A^{-1}& \;\; -   \sigma_S^{-2}\,A_\alpha Q_1 B+B_\alpha Q_2 \\
        B_\alpha^T Q_1 -D_\alpha Q_2 B^T A^{-1}& \;\;- \sigma_S^{-2}\, B_\alpha^T Q_1 B+D_\alpha Q_2  \nonumber
    \end{bmatrix}\,,
\end{align}
where
\bea
Q_1&=&(A-B\sigma^{-2}_S \idm B^T)^{-1}\,,
\\
Q_2&=&\sigma^{-2}_S\,( \idm - \sigma^{-2}_S\,B^T A^{-1}B)^{-1}\,.
\eea
To proceed,
we first note some useful relations involving the inverse of matrices
\begin{align}
    (A-B\frac{\idm}{\sigma^2_S} B^T)^{-1} & \approx \left[\idn + \frac{A^{-1}BB^T}{\sigma^2_S}+\oc(1/\sigma^4_S) \right]A^{-1}\,,\\ 
    (D-B^T A^{-1}B)^{-1} & \approx \frac{1}{\sigma^2_S}\left[\idm 
 + \frac{B^T A^{-1} B}{\sigma^2_S}+\oc(1/\sigma^4_S)\right]\,.
\end{align}
Then,
we define $C_{\alpha}C^{-1}\equiv \calm_\alpha$. For computing the trace in the expressions above, we need the diagonal elements of \mbox{$\calm_\alpha\calm_\beta \equiv {C_\alpha C^{-1}C_\beta C^{-1}}$} 
\begin{align}
\label{eq:Trace_block_split}
   \fc_{\alpha\beta} = \frac{1}{2}\Tr[\calm_\alpha\calm_\beta ] =& \frac{1}{2} \Bigg\{ \underbrace{\Tr{[\calm_{\alpha,11}\calm_{\beta,11}}]}_{\oc(1)} + \underbrace{\Tr{[\calm_{\alpha,12}\calm_{\beta,21}]}}_{\oc(1/\sigma^2_S)} \\ &+ \underbrace{\Tr{[\calm_{\alpha,21}\calm_{\beta,12}]}}_{\oc(1/\sigma^2_S)} + \underbrace{\Tr{[\calm_{\alpha,22}\calm_{\beta,22}]}}_{\oc(1/\sigma^4_S)} \Bigg\} \nonumber\,.
\end{align}
In the above equation the, term in the right-hand-side of the first line denotes the upper left $N\times N$ block of $\calm_\alpha\calm_\beta$ and the term in the second line denotes the lower right $M\times M$ block (with $N$ and $M$ controlling
respectively the pulsar and star numbers). We explicitly indicate
the order of the expansion
in the small $1/\sigma_S^2$
parameter.
\\

In fact,
working in a noise-dominated, large $\sigma_S$
regime for astrometry, we write -- up
to first order in $\sigma_S^{-2}$:
\bea
{\cal M}_\alpha\,=\,{\cal M}_\alpha^{(0)}
+\sigma_S^{-2}\,{\cal M}_\alpha^{(1)}\,,
\eea
with
\bea
{\cal M}_\alpha^{(0)}&=&
\begin{bmatrix}
A_\alpha\,A^{-1}&\,\,0\\
B_\alpha^T\,A^{-1}&\,\,0
\end{bmatrix}\,,
\\
{\cal M}_\alpha^{(1)}&=&
\begin{bmatrix}
A_\alpha\,A^{-1}\,B B^T A^{-1}-B_\alpha B^T A^{-1}&\,\,\,\,B_\alpha-A_\alpha A^{-1} B\\
B^T_\alpha\,A^{-1}\,B B^T A^{-1}-D_\alpha B^T A^{-1}&\,\,\,\,D_\alpha-B^T_\alpha A^{-1} B
\end{bmatrix}\,.
\eea
Explicitly, an expansion of the 
Fisher matrix up to order
$1/\sigma_S^2$ gives the formula
\bea
{\cal F}_{\alpha\beta}
&=&
\frac12 \Tr[{\cal M}_\alpha^{(0)}
{\cal M}_\beta^{(0)}
] +\frac{1}{2\,\sigma_S^2} \Tr[{\cal M}_\alpha^{(0)}
{\cal M}_\beta^{(1)}+{\cal M}_\alpha^{(1)}
{\cal M}_\beta^{(0)}]
\\&=& 
\frac{1}{2}\Tr[A_\alpha \,A^{-1} A_\beta \,A^{-1}] 
\nonumber
\\
&+&
\frac{1}{2\,\sigma_S^2} \Tr[
A_\alpha\,A^{-1} A_\beta A^{-1} B B^T A^{-1}-{A_\alpha A^{-1}} {B_\beta B^T A^{-1}}-{A_\alpha A^{-1}B} {B^T_\beta A^{-1}}]
\nonumber
\\
&+&
\frac{1}{2\,\sigma_S^2} \Tr[
{A_\alpha A^{-1}  B B^T A^{-1}} {A_\beta  A^{-1}}-{B_\alpha B^T A^{-1}}{A_\beta A^{-1}} -{B^T_\alpha  A^{-1}} {A_\beta A^{-1}B} ]
\nonumber
\\
&&+\frac{1}{2\,\sigma_S^2} \Tr[
{B^T_\alpha A^{-1}} {B_\beta } +{B_\alpha } {B^T_\beta A^{-1}}]\,.
\label{exp_FMf}
\eea

\noindent At zeroth order in $1/\sigma^2_S$, we recover the PTA only Fisher matrix -- first
line of eq. \eqref{exp_FMf}.
 At order $\oc(1/\sigma^2_S)$ we obtain the first corrections to the PTA-only forecasts, associated
 with synergies with astrometry. Since the relevant terms always include the matrix $B$,  such corrections include the  pulsar-star correlations.
Notice that this perturbative expansion of the Fisher matrix greatly simplifies the numerical calculations in the astrometric noise dominated regime, since we  only have to deal with matrices of size $N\times N$ and $N\times M $, but not $M\times M$ where $M$ can be of the order $10^5$ or larger. (Recall that $N$
is the number of pulsars, $M$ is twice the number of stars which are monitored.) On the other hand for typical pulsar datasets, currently and in the near future we  have to handle matrices of  size  $N~\sim \oc(100)$, which is 
 manageable with standard
 computing resources. \\

\begin{figure}
    \centering
    \includegraphics[width=0.48\linewidth]{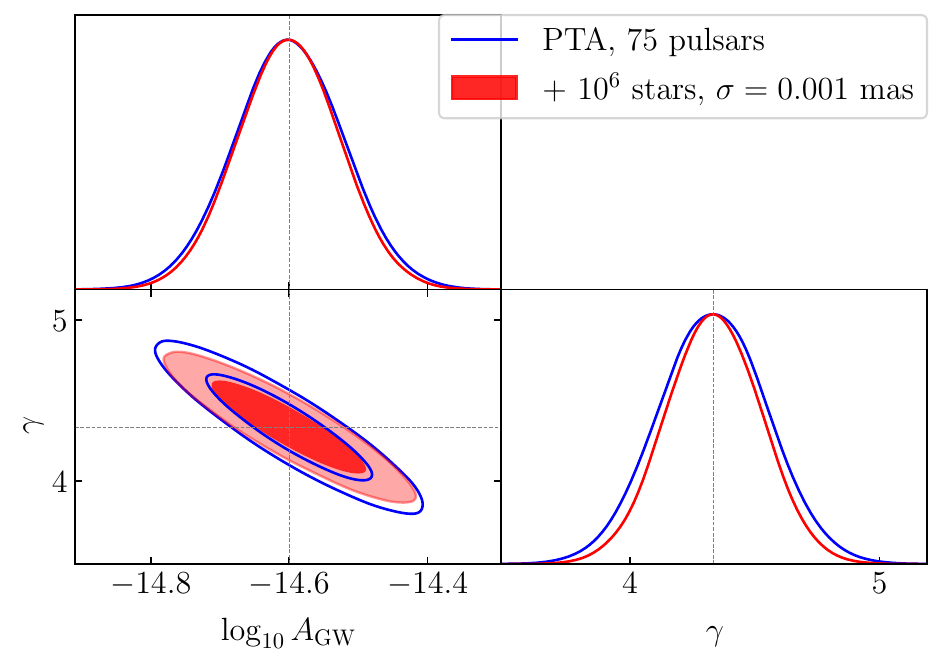}
    \includegraphics[width=0.48\linewidth]{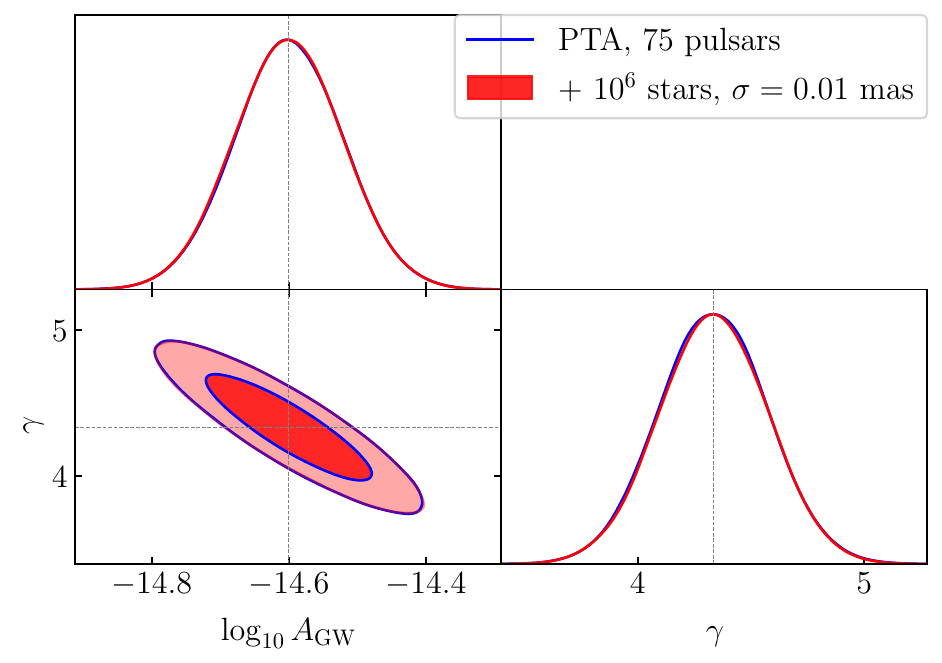}
    \caption{\small Fisher forecast for the amplitude and slope of the SGWB for a fiducial value $\log_{10} A_{\rm GW} = -14.6,\, \gamma = 13/3$. In red, the PTA only case with the number of pulsars and noise parameters chosen to  match the sensitivity of the combined IPTA dataset~\cite{InternationalPulsarTimingArray:2023mzf}. In blue, the forecasts on the addition of PTA--Astrometry cross-correlations with $10^6$ stars. {\bf Left}: astrometry noise $\sigma = 0.001$ mas. {\bf Right}: astrometry  noise $\sigma = 0.01$ mas. }
    \label{fig:Fisher_perturb}
\end{figure}

\begin{table}
    \centering
    \begin{tabular}{|c|c|c|}
    \hline
        & $\boldsymbol{\log A_{\rm GW} }$ & $\boldsymbol{\gamma}$  \\
        \hline
         \textbf{PTA only} & $-14.6 \pm 0.079$ & $13/3 \pm 0.24$ \\
         \hline
         \textbf{PTA + Astrometry, {\boldmath$ \sigma = 0.001$} mas} & $-14.6 \pm 0.072$ & $13/3 \pm 0.19$ \\
         \hline
        \textbf{PTA + Astrometry, {\boldmath$ \sigma = 0.01$} mas} & $-14.6 \pm 0.079$ & $13/3 \pm 0.24$ \\
         \hline
    \end{tabular}
    \caption{\small Fiducial parameter values and marginalised $1\sigma$ limits from the Fisher forecast. We focus on synergy measurements of the SGWB monopole as discussed in section \ref{sec_methhand}.}
    \label{tab:fisher_params}
\end{table}

\noindent We now use
this approach to compute the full Fisher matrix,  combining  the individual per-frequency Fisher matrices. In Fig. \ref{fig:Fisher_perturb}, we present our results for a PTA setup with 75 pulsars and an astrometric setup with $10^6$ stars. The astrometric setup has  {$\Delta\theta_{\rm rms}=0.01,\,0.001\,\rm mas$} and $T_{\rm cad}=\mathrm{year}/15$, i.e. roughly corresponding to the observational cadence of Gaia. Our results indicate that for the case with {$\sigma = 0.001$} mas, we obtain roughly a $\mathcal{O}(10\%)$ improvement over the PTA-only constraints, while for {$\sigma \geq 0.01$} mas, the improvement will be negligible.\footnote{We note that for the {$\sigma=0.001$} mas case, the weak signal limit does not hold for the first few frequency bins where the signal and noise are comparable in magnitude. However, {we have checked (with lower number of stars) that for similar noise vs signal strengths, the results from the series expansion match fairly well with the full numerical results. }} See
Table \ref{tab:fisher_params} for the exact numbers.

\subsubsection{The case of
dipolar anisotropy}
\label{sec_syndip}

We conclude our analysis with a
discussion of the prospects
to detect a possible SGWB dipolar anisotropy through the  synergy
of astrometry and PTA. We aim to quantify the minimal level of anisotropy detectable
by perspective PTA-only data, and with the joint system astrometry-PTA using the same configurations used in the previous section. Although our analysis is tailored to the  kinematic dipole (see eq. \eqref{kinan2}), we expect that similar considerations hold for the case of dipolar statistical anisotropies, of which the kinematic dipole is just a specific case. Given the crucial role that anisotropies may play in distinguishing astrophysical versus cosmological sources of SGWB, such a question is important to address.  Notice that  current upper limits on the magnitude of SGWB anisotropies  lie at $10\%$ level relative to monopole~\cite{NANOGrav:2023tcn,Cruz:2024svc}.

We carry on the analysis with the same approach developed in section \ref{sec:PTA_AstrometryForecast}.
The components $A$, $B$, $D$ entering in the covariance
matrix \eqref{mat_corra} now read
%
\begin{eqnarray}    \label{eq:auto_cross_block_dipole}
    A_{pq} &=& \frac{(\gamma_{pq}^{(0)}  + \gamma_{pq}^{(1)}) I_f}{(4\pi f)^2} + \sigma^2_p \delta_{pq}\,,
    \\
    B_{pa} &=& \frac{(K_{p,a}^{(0)}+K_{p,a}^{(1)})I_f}{{(4\pi)^2 }f}\,,\\
     D_{ab}& =& \frac{(H_{ab}^{(0)}+H_{ab}^{(1)})}{{4\pi}}I_f + \delta_{ab}\sigma^2_{a}\,.
\end{eqnarray}
Besides the monopole,
we now include the contributions
to the dipole ORF: $\gamma^{(1)}$ for PTA \cite{Tasinato:2023zcg}, $K^{(1)}$ for the cross-correlation and astrometry-only $H^{(1)}$ (see section \ref{sec:theory_ORF}). The analysis proceeds in the same manner as the previous section and the results are plotted in Fig. \ref{fig:cross_dipole_forecasts}. The fiducial parameter means and the marginalised $1\sigma$ Fisher errors are collected in Table  \ref{tab:dipole_params}.  

\begin{figure}[t!]
    \centering
    \centering
    \includegraphics[width=0.49\linewidth]{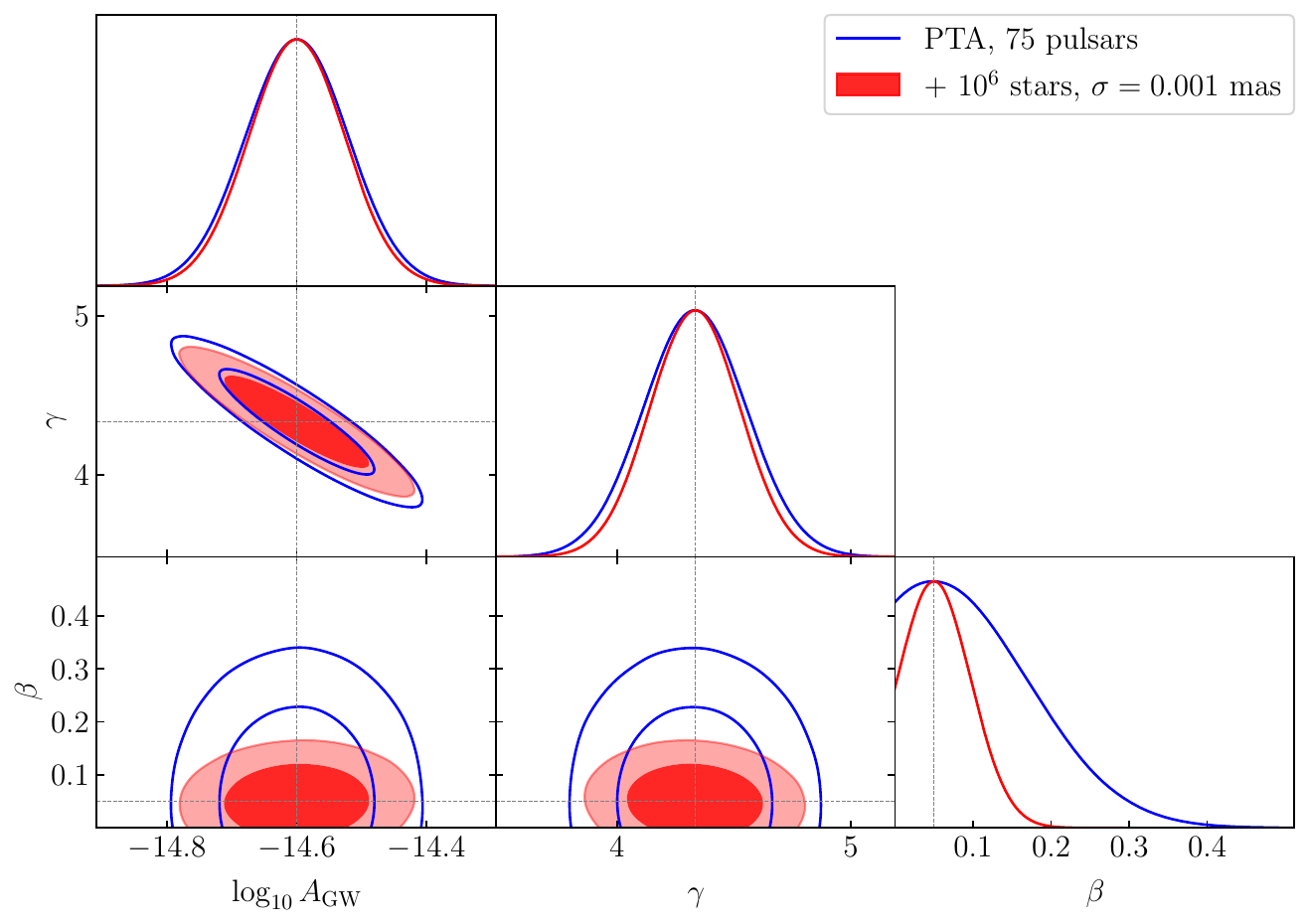}
    \includegraphics[width=0.49\linewidth]{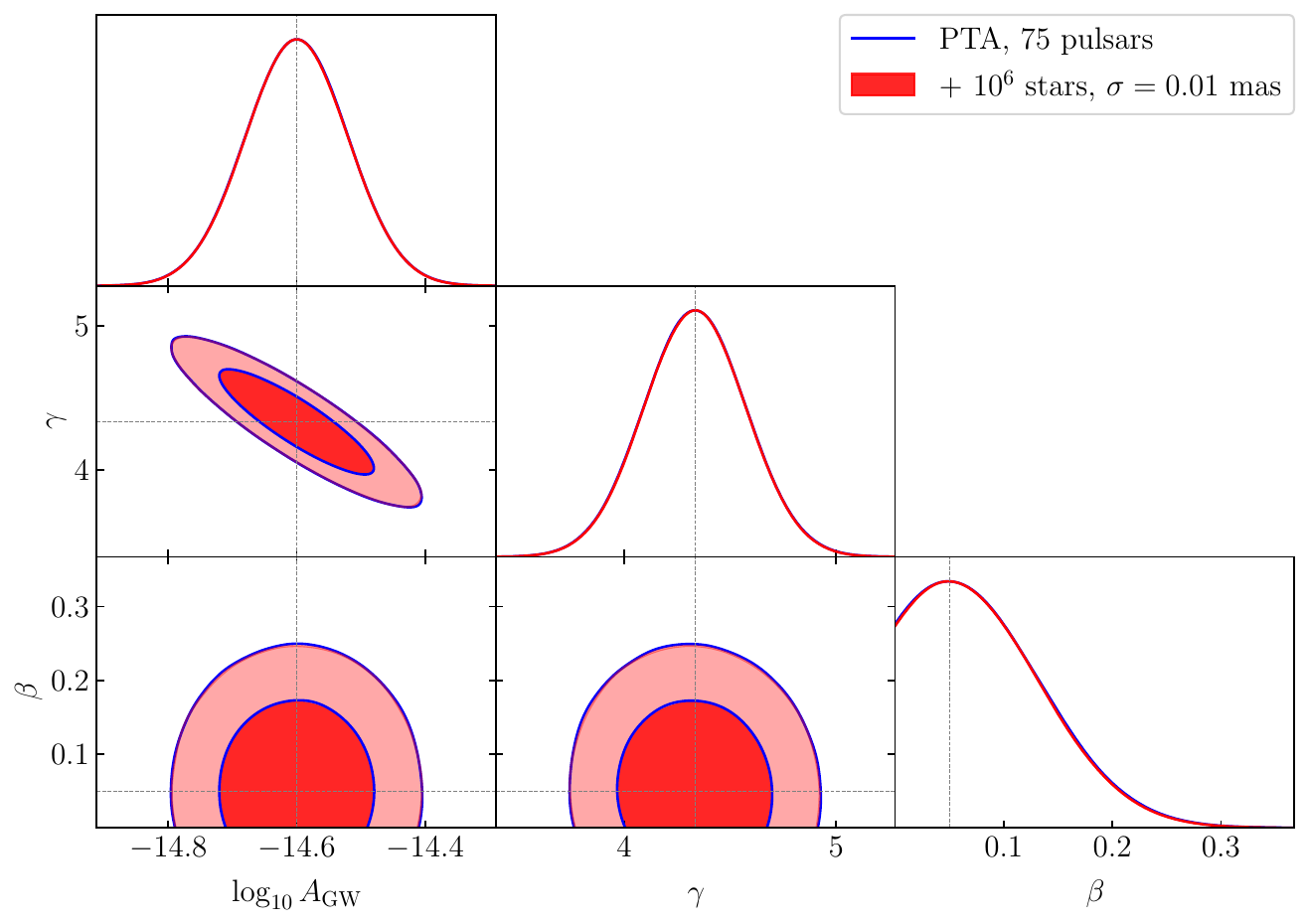}
    \caption{\small Fisher forecast for the SGWB parameters,  with the addition of the dipole.}
    \label{fig:cross_dipole_forecasts}
\end{figure}

\begin{table}[h!]
    \centering
    \begin{tabular}{|c|c|c|c|}
    \hline
        & $\boldsymbol{\log A_{\rm GW} }$ & $\boldsymbol{\gamma}$ & $\boldsymbol{\beta}$  \\
        \hline
         \textbf{PTA only} & $-14.6 \pm 0.079$ & $13/3 \pm 0.24$ & $0.05 \pm 0.081 $ \\
         \hline
        \textbf{PTA + Astrometry, {\boldmath$ \sigma = 0.001$} mas} & $-14.6 \pm 0.072$ & $13/3 \pm 0.18$ & $0.05 \pm 0.048 $ \\        
        \hline
        \textbf{PTA + Astrometry, {\boldmath$ \sigma = 0.01$} mas} & $-14.6 \pm 0.079$ & $13/3 \pm 0.24$ & $0.05 \pm 0.080 $ \\
         \hline
    \end{tabular}
    \caption{\small Fiducial parameter values and marginalised $1\sigma$ limits from the Fisher forecast. In this Table we report limits on the size of dipolar anisotropy, as discussed in section \ref{sec_syndip}.}
    \label{tab:dipole_params}
\end{table}
We  choose the same fiducial $A_{\rm GW}$ and $\gamma$ as in  section \ref{sec_conex}. We select  a level of dipole anisotropy $\beta=5\times 10^{-2}$, finding this value   to be the minimum level of kinematic dipole that will detectable with our PTA + Astrometry setup. Thus, astrometry can again help to tighten the constraints on the dipole anisotropy, which can be useful in determining the origin of the SGWB. 
 {Given that the sensitivity to the SGWB is likely to be dominated by PTA datasets even in the near future, one would likely require futuristic PTA experiments combined with futuristic astrometric surveys to reach the CMB kinematic dipole level. } Reaching the {CMB} kinematic dipole level ({$\beta=1.23\times 10^{-3}$}) with PTA alone requires a substantial increase in the number of pulsars ($N>1000$)~\cite{Cruz:2024svc,Depta:2024ykq} which may only be achievable with SKA. It would be also interesting to explore how cross-correlations with astrometric datasets could work to improve SGWB sensitivity in the SKA-era.

\section{Conclusions}
\label{sec:conclusions}
In the next few years, evidence for the SGWB in the nHz range is likely  to grow as PTA experiments collect more and more data. We can expect
 an increased statistical significance for the Hellings-Downs correlation, as well as tighter constraints on the amplitude and frequency spectrum of the SGWB. The next generation of radio telescopes -- SKA is expected to begin operations in the early 2030s and is projected to observe a large number of pulsars with unprecedented timing precision~\cite{Janssen:2014dka}, possibly leading to precise determination of the SGWB parameters. A complementary probe of the nHz SGWB is provided by astrometry -- corresponding to  the precise monitoring of the positions of a large number of stars. Data from astrometric surveys such as Gaia, has already been used to constrain SGWB in the frequency range $f \lesssim 10^{-9}$ Hz~\cite{Darling:2018hmc,Aoyama:2021xhj,Jaraba:2023djs,darling2024newapproachlowfrequency}. 
 In this work, we have considered how astrometry data in the nHz band will complement PTA observations, and how it can be used to characterize the properties of SGWB beyond what can obtained
 from PTA-only data.

In \cref{sec:theory_ORF}, we first reviewed the theory behind the astrometric detection of SGWB and derived {for the first time} fully covariant, analytical expressions for the kinematic dipole ORFs of the auto-correlation of the astrometric deflections, as well as their cross-correlation with pulsar timing residuals. Our expressions allowed us to manifestly  visualise the sensitivity of the astrometric and PTA setups to the SGWB
 properties, as a function of the locations of monitored objects.

In \cref{sec:forecasts} we studied the sensitivity of the astrometric and PTA setups to the SGWB, focusing on the measurement of the SGWB amplitude, spectral tilt and the magnitude of the kinematic dipole anisotropy. In \cref{sec:Astrometry_estimator} we 
derived  optimal estimators for  SGWB monopole and dipole measurements in this context,  built in terms of quadratic combinations of the astrometric deflections. We  used these formulas to forecast the sensitivity of astrometry to the SGWB monopole and dipole, assuming a large number of stars uniformly distributed  across the sky.  In~\cref{sec:PTA_AstrometryForecast}, we analysed the joint PTA-Astrometry setup and showed that cross-correlating PTA and Astrometry data could tighten the constraints on SGWB obtained from PTA data alone. 
We used Fisher forecasts to calculate the sensitivity to the SGWB amplitude, spectral shape and dipole, finding that an astrometric survey with $0.01$ mas astrometric precision and the typical number of sources and cadence of Gaia could lead to noticeable improvements over current PTA only SGWB constraints. Improvements on the PTA constraints can be quite useful, since tighter constraints on the SGWB parameters can be used to rule out models and potentially distinguish between and astrophysical or cosmological signal.  
 {Reaching the level of sensitivity required to detect the CMB level kinematic dipole anisotropy in the SGWB is likely to require both futuristic PTA experiments such as SKA, as well as futurustic  astrometric surveys.}

The flood of upcoming data from PTA and astrometry experiments presents both challenges and opportunities. The possibility of a joint PTA-astrometry data analysis will hold exciting potential since cross-correlations between timing residuals and astrometric deflections could be leveraged to deliver tighter SGWB constraints compared to either of the experiments alone. Thus, developing efficient implementations of joint PTA-astrometry analysis will be crucial for harnessing the power of these cross-correlations. It would also be interesting to further explore such synergies in the context of general SGWB anisotropies and study their detectability. We leave such pursuits to future work.
 
\subsection*{Acknowledgements}
We are partially funded by the STFC grants ST/T000813/1 and ST/X000648/1. 
We also acknowledge the support of the Supercomputing Wales project, which is part-funded by the European Regional Development Fund (ERDF) via Welsh Government. We acknowledge the use of the \texttt{GetDist} software package for visualising the results of the Fisher forecasts~\cite{Lewis:2019xzd}. {The codes used to perform the numerical calculations and plot the results in this paper are publicly available on \href{https://github.com/MarisolJC/Astrometry_meets_PTAs}{GitHub}}.  For the purpose of open access, the authors have applied a Creative Commons Attribution licence to any Author Accepted Manuscript version arising.

\begin{appendix}

\section{Spectrum of angular deflection fluctuations}
\label{sec_spec}
In this Appendix we make use
of our analytical, covariant
expressions
for the ORF to show that dipolar
anisotropies induce correlations
between electric and magnetic
components of star deflections. We 
use the notation of \cite{Book:2010pf}, to which we refer the reader for more details on the nomenclature.

The $EB$ correlation will be given by 
\be
C_{Elm\,Bl'm'}=\frac{1}{\ell(\ell+1)}\int{
d^2\Omega_n d^2\Omega_{n'}Y^*_{lm}(n)Y_{l'm'}(n') \beta(4-n_{\Omega})\beta^{EB}
\,,
}
\ee
where 
\be
 \beta^{EB}
 = \nabla_i\nabla_b'
\left[\varepsilon_{jab}n_a H_{ij}^{(1)}\right]\,.
\ee
Where $\nabla_i$ and $\nabla_b'$ are normal 3D derivatives with respect to ${\bf z}$ and ${\bf z'}$ with   ${\bf n =z/|z|}$, ${\bf q =z'/|z'|}$. 
 Now, we only need to focus on the parts of $H_{ij}^{(1)}$ (see \eqref{eq:H1ij}), which are not invariant under ${\bf n\to -n}$ and ${\bf q\to -q}$, as these give a zero contribution \cite{Book:2010pf}. Thus, we only have the following contribution: 
\begin{equation}\label{eq:betaEB}
\begin{split}
\beta^{EB}&=  \nabla_i\nabla_b'
\left[a_1 \epsilon_{jab} n_a\left( \epsilon^{ilm}n_l q_m [{\bf(n\cdot q)}q^j-n^j]+\epsilon^{jlm}n_lq_m[{\bf(n\cdot q)}n^i-q^i]\right)\right] \\
&=  \nabla_i\nabla_b' \left[a_1 T^{bi} \right]  \,,
\end{split}
\end{equation}
where
\be
T^{bi} =  (q^b-({\bf n\cdot q}) n^b)(q^i +n^i)(({\bf n\cdot q}) -1) - (n^i q^b- ({\bf n\cdot q}) \delta^{ib}) (({\bf n\cdot q})^2-1)\,,
 \ee
and  $a_1$  is defined in \eqref{eq:defa1}. 
So we need $\nabla_i X, \nabla_b' X, \nabla_i\nabla_b' X$, where $X= a_1, T^{bi}$. We can use the relations 
$\nabla_i n_j =\delta_{ij} - n_i n_j$, $\nabla_i' q_j =\delta_{ij} - q_i q_j$, $\nabla_i q_j=\nabla_i' n_j =\nabla_i v_j=\nabla_i' v_j =0$, as well as
\be
\nabla_i({\bf n\cdot q}) = q_i-({\bf n\cdot q})n_i\,,\qquad \nabla_b'({\bf n\cdot q}) = n_b-({\bf n\cdot q})q_b\,.
\ee
Using these, we find:
\begin{subequations}\label{eq:DgammaT}
 \begin{align}
 \nabla_i\nabla_b'(a_1) \,T^{bi} = & -\frac{\pi}{12}\varepsilon^{ljk}v_l n_j q_k \, F_1 \left[2-\left(-3+\frac{F_{1,y}}{F_1}(\bn \cdot \bq-1)-2\,\bn \cdot \bq\right)(\bn \cdot \bq)^2(\bn \cdot \bq-1)\right] \,, \\
 \nabla_i(a_1)\nabla_b'(T^{bi}) = &-\frac{\pi}{12}\varepsilon^{ljk}v_l n_j q_k \,F_1 \,(1-\bn \cdot \bq)[-1+(\bn \cdot \bq)^2(\bn \cdot \bq(\bn \cdot \bq-7)-1)] \,, \\
\nabla_b'(a_1)\nabla_i(T^{bi}) = & -\frac{\pi}{6}\varepsilon^{ljk}v_l n_j q_k \,(1+\bn \cdot \bq+(\bn \cdot \bq)^2(2+\bn \cdot \bq))\left(F_2\, \bn \cdot \bq+\frac{F_1}{2}(1-(\bn \cdot \bq)^2)\right) \,, \\
(a_1)\,\nabla_i\nabla_b'(T^{bi}) =& \frac{\pi}{6}\varepsilon^{ljk}v_l n_j q_k\,F_2\left[1-\bn \cdot \bq(2+\bn \cdot \bq[4+\bn \cdot \bq(3\bn \cdot \bq-10)])\right],
 \end{align}
\end{subequations}
where we defined 
\bea
F_1&=& \frac{(-1 + 4 y - 14 y^2 + 11 y^3 - 3 y^2 (2 y + 1) \ln(y))}{(y^2 (1 - y)^4)} \,,\\
F_2&=&\frac{\left(1 - 4 y - \frac{3 y^2 \ln(y)}{(1 - y)}\right)}{y (1 - y)^2}\,,
\eea
to simplify notation and $F_{1,y}=dF_1/dy$. Combining and simplifying  these, we get for 
\eqref{eq:betaEB}:

\bea\label{eq:betaEBfull}
\beta^{EB} = \frac{\pi}{12}\,{\bf v\cdot(n\times q)} G(y)\,,
\eea
where 
\begin{equation}
\begin{split}
G(y)&=-\frac{2}{y^2(y-1)^5} \Bigg[ (y-1) (-1 + 6 y - 42 y^2 + 170 y^3 - 481 y^4 + 800 y^5 - 684 y^6 + 
   208 y^7 + 96 y^8)  \\
&  \hskip3cm -3 y^2 (-1 - y + 13 y^2 - 77 y^3 + 198 y^4 - 236 y^5 + 128 y^6) \ln(y) \Bigg]\,,
\end{split}
\end{equation}
where $y$ is defined in eq.~\eqref{defoy}. Note that $\beta^{EB}$ depends on the angles between ${\bf v}$ with ${\bf n}$ and ${\bf q}$ as well as $\zeta$.

\section{Angular dependence of overlap reduction functions}
\label{Trace_Appendix}
In this Appendix we report explicit
formulas for the traces
of matrix combinations used in section \ref{sec:theory_ORF} for representing
the sensitivity of the PTA and astrometry system to the properties
of SGWB. 
From equations \eqref{eq:H0ij} and \eqref{eq:K0i}
\bea
\Tr[\hb_{0}\hb_{0}] & = & \frac{\pi ^2 \left(\cos ^2(\zeta )+1\right)}{9 (\cos (\zeta
   )+1)^2} \times \label{TrH0H0} \\
   & & \left(-7 \cos ^2(\zeta )-2
   \cos (\zeta )+6 (\cos (\zeta )-1)^2 \ln \left(\sin
   ^2\left(\frac{\zeta }{2}\right)\right)+5\right)^2 \,;\nonumber \\
\mathbf{K}_{0}\mathbf{K}_{0}^{T}& = & \frac{16}{9} \pi^2 \tan^2 \left(\frac{\zeta_{sp} }{2}\right) \times \label{eq_K0K0} \\
& & \left(-3 \ln
   (1-\cos (\zeta_{sp} ))+\cos (\zeta_{sp} ) \left(3 \ln \left(\sin
   ^2\left(\frac{\zeta_{sp} }{2}\right)\right)-2\right)-2+\ln (8)\right)^2, \nonumber
\eea
where $\zeta$ is the angle between the stars at directions $\bn$ and $\bq$, while $\zeta_{sp}$ is the angle between a star in direction $\bn$ and a pulsar at direction $\bx$.

With equations \eqref{eq:H1ij} and \eqref{eq:K1i} we get
\bea
\Tr[\hb_{1}\hb_{1}] & = &  \frac{64 \pi ^2 \left(\alpha_{H}  (Av)^2 \sqrt{1-y} y^2+4 \delta_{H}^2
   (y-1) y^4 \left(\beta_{H} -\gamma_{H}  \sqrt{1-y}\right)\right)}{9
   \left((nq)^2-1\right)^2 (1-y)^{5/2}} \,;\label{TrH1H1} \\
\mathbf{K}_{1}\mathbf{K}_{1}^{T} & = & \frac{4 \pi ^2 \left(nx^2 -1\right) \left((nx +1)^2
   \left(\alpha_K  \beta_K ^2-(A_{1}v)^4 (5-6 nx)^2\right)+12
   \gamma_K  \ln \left(\frac{1-nx}{2}\right)\right)}{9
   (nx+1)^4 ((A_{1}v)-(A_{1}v)nx )^2}, \, \, \, \, \, \, \, \, \, \label{eq_K1K1}
\eea
where
\bea
\alpha_H & = & -(y-1)^2 \left(nq^2 y^2 (2 y+1)^2+2 nq (1-4 y)^2+y^2 (2
   y+1)^2\right)+ \\ 
   & & 6 (y-1) y^2 \left(2 \left(nq^2+1\right) 
   y^2+nq (nq+8) y-2 nq+y\right) \ln (y)-9
   (nq+1)^2 y^4 \ln ^2(y) \,,\nonumber \\
\beta_H & = & (nq)^2 (nv) (y-1) \sqrt{\frac{(Av)^2}{y-1}+4
   \left((nv)^2-1\right) y} \\
   & & -(nv) \sqrt{1-y} \sqrt{4
   \left((nv)^2-1\right) (y-1) y-(Av)^2}\,, \nonumber \\
\gamma_H & = & (nv)^2 \left((nq)^2-2 y+1\right)+(nq)^2 (-y)+y \,, \\
\delta_{H} & = & 1 + y - 2 y^2 + 3 y \ln{y}\,,
\eea
and
\bea
\alpha_K & = & -1 + (nv)^2 + (nx)^2 - 2nv(nx)(vx) + (vx)^2 \,, \\
\beta_K & = & nv (4 + nx (-7 + 2 nx)) + (4 + 3 (-2 + nx) nx) vx \,, \\
\gamma_K & = & (-1 + nx)^2 (1 + nx) ((A_{1}v)^4 (-5 + 6 nx) + (nv + vx) \alpha_K \beta_K) + \nonumber \\ 
& & 3 (-1 + nx)^4 (-(A_{1}v)^4 + (nv + vx)^2 \alpha_K) \ln{\left(\frac{1 - nx}{2}\right)}.
\eea

When the vector $\bn$ is parallel to $\bv$, then $Av=0$ and equation \eqref{TrH1H1} is simpler. For this case $\mathbf{K}_{1}\mathbf{K}_{1}^{T}$ is calculated with \eqref{K1i_specialC}. More specifically when $\bn = -\bv$, we get
\bea
\Tr[\hb_{1}\hb_{1}] & = & \frac{4 \pi ^2 (\cos (\zeta )-1)^2 \left(\cos ^2(\zeta )+1\right)
   }{9 (\cos
   (\zeta )+1)^2} \times \\
& & \left(\cos ^2(\zeta )-\cos (\zeta )-\ln (8) \cos (\zeta )+3 (\cos
   (\zeta )-1) \ln (1-\cos (\zeta ))-2+\ln (8)\right)^2 \,, \nonumber \\
\mathbf{K}_{1}\mathbf{K}_{1}^{T} & = & \frac{1}{9} \pi ^2 \tan ^2 \left(\frac{\zeta_{sp} }{2}\right) \times \\ & & \left(2 \cos
   (\zeta_{sp} )-3 \cos (2 \zeta_{sp} )-12 (\cos (\zeta_{sp} )-1) \ln \left(\sin^2 \left(\frac{\zeta_{sp} }{2}\right)\right)+5\right)^2. \nonumber 
\eea
Similar expressions for the case $\bn = \bv$.

\section{Optimal Estimators}
\label{sec:Astrometry_estimator}
{
In this section we briefly discuss how to build optimal estimators for the coefficients $p_n$ 
formally appearing
in the sum~\eqref{est2ddef}, later specialising to the monopole and  dipole case with $p_0$ and $p_1$. The determination of these coefficients allows us to
infer the properties of the SGWB.
First, we recall here that we can expand the astrometric deflection correlation in terms of these basis coefficients $p_n$ as 
\begin{align}
    \qev{\delta n_a^i \, \delta n_b^j} \equiv C^{ij}_{ab} = \sum_{n} p_n H_{ab,n}^{ij} + N_{ab}^{ij}\,,
\end{align}
in terms of the response $H$ and the noise $N$. Now, following the approach previously developed for the CMB~\cite{Tegmark:1996qt}, a quadratic estimator for $p_n$ can be formally expressed as
\begin{align}
\label{eq:def_pn_hat}
    \phat_n = \delta n^i_a E_{ab,n}^{ij} \delta n^j_b - b_n\,,
\end{align}
in terms of a symmetric matrix $E_{ab,n}$ and a vector $b_n$. 

We wish to determine the corresponding  values of these quantities which render
the estimator unbiased i.e. $\qev{\phat_n} = p_n$, and minimize its variance. Taking the expectation value of the estimator we find
\begin{align}
    \qev{\phat_n} = \sum_{n'}W_{nn'}p_{n'} + \Tr[\eb_n \nb] - b_n,\hskip0.5cm {\rm with} \hskip0.5cm W_{nn'} \equiv \Tr[\hb_{n'}\eb_n]\,.
\end{align}
Thus, the  quantity $b_n$ should be chosen as $b_n = \Tr[\eb_n \nb]$
to ensure that the estimator is  unbiased.  The covariance of matrix of the estimators $\mathrm{Cov}(\phat_n,\phat_{n'}) \equiv \qev{ (\phat_n - \qev{\phat_n}) (\phat_{n'} - \qev{\phat_{n'})} }$, assuming that the astrometric deflections obey a Gaussian distribution, can be written as 
\begin{align}
\mathrm{Cov}(\phat_n,\phat_{n'}) &= \left[C_{ad}^{il}C_{bc}^{jk}+C_{ac}^{ik}C_{bd}^{kl} \right]E^{ij}_{ab,n}E^{kl}_{cd,n'} \\
&=2\Tr[\cb \eb_n \cb \eb_{n'}]\,.
\end{align}
The solution for the matrix $\eb_n$  which minimises this variance under the normalisation constraint $W_{nn} = 1$ is given by~\cite{Tegmark:1996qt}
\begin{align}
    \eb_n = \frac{\cb^{-1} \hb_n\cb^{-1}}{\Tr[\cb^{-1}\hb_n \cb^{-1}\hb_n ]}\,,
\end{align}
so that 
\begin{align} W_{nn'} 
    = \frac{\Tr\left[\cb^{-1} \hb_{n'} \cb^{-1} \hb_n \right]}{\Tr\left[\cb^{-1}\hb_n \cb^{-1}\hb_n \right]}\,.
\end{align}
The variance of the individual $\phat_n$ is thus
\begin{align}
    {\text{Var}} (\phat_n)  = \frac{2}{\Tr\left[ \cb^{-1}\hb_n \cb^{-1}\hb_n   \right] }\,.
\end{align}
This estimator is optimal in the sense that it is unbiased and that its variance is equal to the what we obtained using the Fisher matrix approach in \cref{sec:Fisher_astro_only}, making it the minimal variance estimator that saturates the Cramer-Rao bound (see~\cite{Tegmark:1996bz,Tegmark:1996qt} for a detailed discussion).

The expressions obtained here are completely general and valid for arbitrary distributions of stars with arbitrary noise properties which are encoded in the noise matrix $\boldsymbol{N}$. Further simplifications can be made in the noise dominated limit with the diagonal noise where the solution reduces to  
\begin{align}
    \eb_n = \frac{\hb_n}{\Tr[\hb_n \hb_n]},\quad 
W_{nn'} = \frac{\Tr[\hb_n \hb_{n'}]}{\Tr[\hb_n \hb_n]}\,.
\end{align}
Finally, in the special case where one only wishes to estimate a single $p_n$, e.g. the monopole, from the data, the estimators become trivial
\begin{align}
    \hat p_n = \frac{\delta n^i_a H_{ab,n}^{ij} \delta n_b^j}{(\Tr[\hb_{n}\hb_{n}])}\, - b_n.
\end{align}
}


\end{appendix}
{\small
\addcontentsline{toc}{section}{References}
\bibliographystyle{utphys}
\bibliography{astrometry}
}

\end{document}